\documentclass[10pt]{article}

\usepackage{amsmath}
\usepackage{amssymb}

\usepackage{graphicx}

\usepackage{cite}

\usepackage{color} 

\usepackage{amssymb,amsfonts,amsmath}
\usepackage{dcolumn}% Align table columns on decimal point
\usepackage{multirow}
\usepackage{longtable}
\usepackage[labelfont=bf,labelsep=period,justification=raggedright]{caption}

\makeatletter
\renewcommand{\@biblabel}[1]{\quad#1.}
\makeatother

\date{}

\usepackage{epsfig}
\usepackage{url}
\usepackage{colortbl, pstricks}

\newcommand{\req}[1]{Eq.~(\ref{#1})}
\newcommand{\avg}[1]{\langle #1\rangle}
\newcommand{\fig}[1]{Fig.~\ref{#1}}
\newcommand{\tab}[1]{Table \ref{#1}}

\begin{document}

% Title must be 150 characters or less
\begin{flushleft}
{\Large
\textbf{Leaders in Social Networks, the {\it{Delicious}} Case}
}
% Insert Author names, affiliations and corresponding author email.
\\
Linyuan L\"u$^{1,2,3}$,
Yi-Cheng Zhang$^{1,2,3,\ast}$,
Chi Ho Yeung$^{1,2,3}$,
Tao Zhou$^{1,2,3}$
\\
\bf{1} Research Center for Complex System Science, University of Shanghai for Science and Technology, Shanghai 200093, People's Republic of China
\\
\bf{2} Web Sciences Center, University of Electronic Science and Technology of China, Chengdu 610054, People's Republic of China
\\
\bf{3} Department of Physics, University of Fribourg, Chemin du Mus\'ee 3, CH-1700 Fribourg, Switzerland
\\
$\ast$ E-mail: Corresponding yi-cheng.zhang@unifr.ch
\end{flushleft}

% Please keep the abstract between 250 and 300 words
\section*{Abstract}
Finding pertinent information is not limited to search engines.
Online communities can amplify the influence of a small number of power users
for the benefit of all other users.
Users' information foraging in depth and breadth can be greatly enhanced by choosing suitable 
leaders.
For instance in delicious.com,
users subscribe to leaders' collection
which lead to a deeper and wider reach not achievable with search engines.
To consolidate such collective search,
it is essential to utilize the leadership topology and identify influential users.
Google's PageRank,
as a successful search algorithm in the World Wide Web,
turns out to be less effective in networks of people.
We thus devise an adaptive and parameter-free algorithm,
the LeaderRank,
to quantify user influence.
We show that
LeaderRank outperforms PageRank in terms of ranking effectiveness,
as well as robustness against manipulations and noisy data.
These results suggest that leaders who are aware of their clout may
reinforce the development of social networks,
and thus the power of collective search.

% Please keep the Author Summary between 150 and 200 words
% Use first person. PLoS ONE authors please skip this step. 
% Author Summary not valid for PLoS ONE submissions.   
\section*{Author Summary}

\section*{Introduction}
Many social networks such as \emph{twitter.com} and
\emph{delicious.com} allow millions of users to interact, 
in which some members hold much larger influence than the others.
Identifying these influential users is not easy.
Yet it is essential to identify them in social networks:
what an online community can collectively achieve is to enhance the power of individuals
in discovering new information in depth and breadth that 
no individual can even contemplate,
and an effective way is to make use of influential users.
We take the World Wide Web as an example.
Though many useful pages are out there, the sheer size of WWW 
creates a great barrier for comprehensive information exploration.
Besides search engines which can be of great help, 
there is another mode of information acquisition through leveraging the network power,
getting useful webpages from different experts.
This collective search through social networks \cite{lampe06,vieira07}
may one day complement the current search paradigm which is based on isolated users shooting queries.

Delicious.com, 
previously known as del.icio.us, 
is a representative case that we will focus in this paper. 
Its primary function for individuals is to collect useful bookmarks with tags, 
such that thousands of bookmarks can be easily recalled. 
But for many users,
its new function of networking people is more interesting. 
In delicious.com,
users can select other users to be their \emph{leaders}, in the sense that
the bookmarks of the leaders are often useful 
and subscriptions to these bookmarks will be automatic. 
The subscribers,
which we call \emph{fans},
can in turn be the leaders of other users.
Out of the 7 million users, 
which is still a rapidly increasing number in delicious.com,
about half a million users are linked in a big cluster by these leader-fan relations.
We call this big cluster the \emph{leadership network}. 
Actually this seemingly minority group include the most active users.

Although this leadership network is highly informative for leader identification,
to well utilize the network is challenging \cite{easleyBook,kleinberg99,park05,radicchi09,chen07}.
First of all,
the leadership structure is complex
and going upstream by 
indefinitely climbing up the ladder of leaders is not illuminating.
In addition,
considering only the leaders alone provides no absolute measure of influence,
as it is the entire upstream connection which act as the information sources
and contribute to the influence of a user.
Similarly, 
as we shall see in our experiments,
merely counting the number of fans is not a good way to quantify the leader significance.
A sophisticated model however could reveal the intrinsic structure and identify the worthy leaders.

To well utilize the leadership network
we shall devise a method akin to {\it PageRank} \cite{page99, brin98},
which effectively ranks webpages based on the hyperlink network.
However,
the leadership network is fundamentally different
as personal relationships are quickly evolving,
which makes adaptability essential for ranking users.
For instance,
the probability which describes the random information acquisition 
should self-adjust when users add or remove leaders.
While this probability is governed by an external parameter in PageRank,
we devise our \emph{LeaderRank} algorithm where this probability is adaptive and personalized,
leading to a parameter-free algorithm readily applicable to any type of graph.
This advantage eliminates the frequent needs of parameter tests and calibration
of PageRank on fast evolving networks.
Simulations show that our LeaderRank algorithm outperforms PageRank
in identifying users who lead to quick and wide spreading of useful items.
Moreover,
LeaderRank is more tolerant of noisy data
and robust against manipulations.

In addition to ranking,
the present study may shed light on the future design of community rules
and online social networks.
Leader identification reinforces well-placed individuals 
to go deeper and wider in information exploration,
where the whole society benefits from the collective outputs.
A robust ranking algorithm also discourages people 
from manipulations \cite{masum04}. 
In this paper,
we will compare ranking based on the leadership network
with simple ranking based on the number of fans.
By conducting simulations and experiments,
we will see how ranking algorithms identify influential users in social networks.
Interested readers may try the webpage \url{http://rank.sesamr.com},
where we implement LeaderRank to rank users in delicious.com.

% Results and Discussion can be combined.

\section*{Methods and Materials}

In many online applications,
users are able to select other users to be their sources of information.
We represent these user-user relations by a network 
with directed links pointing from fans to their leaders.
The link direction corresponds to votes from fans for their leaders,
and popular leaders would have a large number of in-links.
We take this convention as it matches the direction of random walk
in our algorithm,
though the direction of information flow in the network is {\em opposite},
i.e. from leaders to fans.
Our aim is to rank all the users based on this network topology. 

\subsection*{LeaderRank}
\begin{figure}[!ht]
\begin{center}
%\includegraphics[width=10cm]{example.eps}
%\leftline{\hspace{3cm}\mbox{(a)}\hspace{6cm}\mbox{(b)}}
\centerline{\epsfig{figure=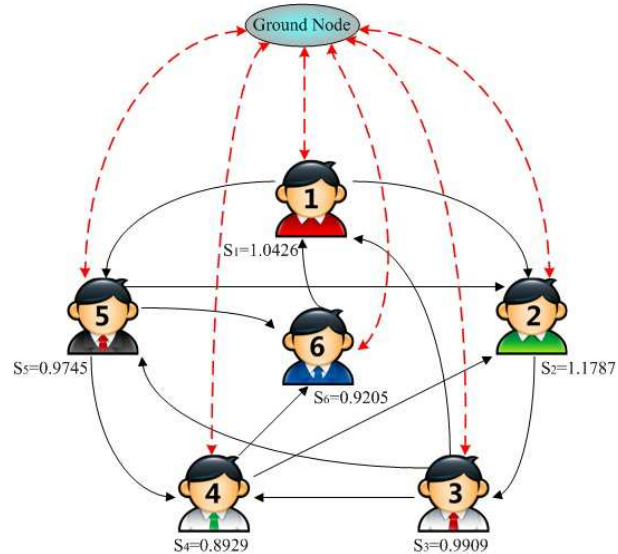, width=0.5\linewidth}}
\caption{An illustration of the ground node and the LeaderRank algorithm. 
The social network consists
of six users and 12 directed links. 
The final ranking scores are labeled next to the corresponding
users.} 
\label{example}
\end{center}
\end{figure}

We consider a network of $N$ nodes and $M$ directed links.
Nodes correspond to users
and links are established according to the relations among leaders and fans.
To rank the users,
we introduce a {\it ground node} which 
connects to every user through bidirectional links
(see Fig.~\ref{example} for an illustration). 
The network thus becomes strongly connected
and consists of $N+1$ nodes and $M+2N$ links.
To start the ranking process,
we assign to each node, except for the ground node, one unit of
resource which is then evenly distributed to the node's neighbors
through the directed links. 
The process continues until steady state is attained.
Mathematically,
this process is equivalent to random walk on the directed network,
and is described by the stochastic matrix $P$
with elements $p_{ij} = a_{ij}/k_i^{\rm out}$
representing the probability that a random walker at $i$ goes
to $j$ in the next step.
$a_{ij}=1$ if node $i$ points to $j$ and 0 otherwise, 
while $k_i^{\rm out}$ denotes the out-degree,
i.e. the number of leaders, of $i$. 
This probability flow thus corresponds to the vote from fan $i$
to leader $j$.
Denoting by $s_i(t)$ the score of node $i$
at time $t$, we have
\begin{equation}
s_i(t+1)=\sum_{j=1}^{N+1}\frac{a_{ji}}{k_j^{\rm out}}s_j(t). \label{equation1}
\end{equation}
The initial scores are given by $s_i(0)=1$ for all node $i$ (other than the ground node)
and $s_{\rm g}(0)=0$ for the ground node.

The presence of the ground node makes $P$ irreducible,
as the network is strongly connected.
The ground node also ensures the co-existence of loops of size 2 and 3 from any node,
which implies $P^6$ is positive,
i.e. all elements of $P^6$ are greater than zero.
As $P^n$ is positive for some natural number $n$,
the non-negative $P$ is primitive.
By the Perron-Frobenius theorem, 
$P$ has the maximum eigenvalue 1 with an unique eigenvector.
We outline the proof of primitivity and convergence in \emph{Supporting Information} (\emph{SI}).
The score $s_i(t)$ for all $i$ thus 
converges to a unique steady state denoted as $s_i(t_c)$,
where $t_c$ is the convergence time.
At the steady state,
we evenly distribute the score of the ground node to all other nodes
to conserve scores on the nodes of interest.
Thus we define the final score of a
user to be the leadership score $S$, namely
\begin{equation}
S_i=s_i(t_c)+\frac{s_{\rm g}(t_c)}{N}, \label{equation2}
\end{equation}
where $s_{\rm g}(t_c)$ is the score of the ground node at steady state.
Based on the above properties,
there are several advantages of applying LeaderRank in ranking,
which include:
(i) parameter-freeness,
(ii) wide applicability to any type of graph,
(iii) convergence to an unique ranking, and
(iv) independence of the initial conditions.

To illustrate the ranking process,
we provide a simple ranking example in Fig.~\ref{example}.
After convergence, 
the final scores of
the six users are $S_1=1.0426$, $S_2=1.1787$, $S_3=0.9909$,
$S_4=0.8929$, $S_5=0.9745$ and $S_6=0.9205$, respectively.
Therefore,  
user 2 is ranked top by the LeaderRank algorithm.

\subsection*{PageRank}
We briefly describe the PageRank algorithm,
with which we compare our ranking results.
PageRank forms the basis of the Google search engine
and represents a random walk on the hyperlink network.
A parameter $c$ is introduced as the probability for a web surfer to
jump to a random website
and $1-c$ is the probability for the web surfer to continue browsing 
through hyperlinks.
$c$ is thus called the {\it return probability},
i.e. the probability that the web surfer returns and starts a new random walk.
In this case,
$s_i(t)$ of a webpage $i$ at time $t$ is given by 
\begin{equation}
s_i(t+1)=c+(1-c)\sum_{j=1}^{N}\left[\frac{a_{ji}}{k_j^{\rm out}}(1-\delta_{k_j^{\rm out},0})
+\frac{1}{N}\delta_{k_j^{\rm out},0}\right]s_j(t). \label{equation3}
\end{equation}
where $\delta_{a,b}=1$ when $a=b$ and 0 otherwise.
The first and second term respectively correspond to the contributions
from random surfers and from surfers arriving through hyperlinks.

Before comparing the ranking results,
there are several drawbacks in applying PageRank to social networks.
Firstly,
return probability is essential in PageRank \cite{page99, brin98}
as algorithmic convergence is only guaranteed on strongly connected networks.
This introduces a parameter to the algorithm,
and results in the frequent need of extensive tests on parameter and evaluation metrics,
which makes PageRank maladaptive to the fast evolving social networks.
In addition,
return probability is identical for all users irrespective of their significance.
For dangling users (those without leaders),
specific treatments are required to distribute all their probability back to the network uniformly \cite{page99}.
All these drawbacks limit the potential of applying PageRank 
to rank users in social networks,
as well as other ranking tasks.

\subsection*{Differences between LeaderRank and PageRank}
An obvious difference between LeaderRank and PageRank lies in the formulation,
where the ground node in LeaderRank plays an important role in regulating probability flows,
making LeaderRank parameter-free.
An essential difference lies in the heart of dynamics,
as in LeaderRank the score flow to the ground node is inversely proportional to the number of selected leaders,
while there is no such relation in PageRank.
Mathematically,
the score flow to the ground node is analogous to the return probability in PageRank,
and the dependence of score flow on the number of leaders makes LeaderRank adaptive to fast evolving networks.
The inverse proportion is reasonable,
as nodes with a small number of leaders receive less information
and hence acquire more information from the ground node
(which corresponds to a larger score flow to the ground node).
The same happens on the Internet,
as web surfers surfing on websites with small out-degree 
have limited choices of hyperlink
and by higher chance jump to another random website.
More detailed discussions are given in the first section of \emph{SI}.

\subsection*{Data description}
We apply the LeaderRank algorithm on the leadership network 
obtained from the world-largest online bookmarking website, 
delicious.com, 
to rank users according to their importance.
Users in delicious.com are allowed to collect URLs as bookmarks, 
and are encouraged to select a list of leaders
as sources of information. 
The dataset we are going to test was collected at May 2008, 
which consists of 582377 users and 1686131 directed links. 
Out of which 571686 users belong to the giant component, 
while the total users in other components are less than $0.1\%$ of the giant component. 
Actually, the numbers of users in the second to fifth largest components are respectively 58, 53, 44 and 35. 
We thus study only the largest component. 
The number of directed links in the largest component is 1675008, 
of which 338756 links (169378 pairs) are reciprocal. 
If the network is considered as an undirected network, 
the clustering coefficient \cite{watts1998}
and assortativity coefficient \cite{newman2002} are respectively 0.241 and -0.012,
while the average shortest distance between users is approximately 5.104.

\section*{Results}

We first show the difference among the rankings obtained by
LeaderRank, PageRank and the number of fans.
Table \ref{tab_diffRank} shows the top 20 users ranked by the three approaches.
%We plot also the ranking by the number of fans as compared to the ranking by LeaderRank in \fig{fig_infr},
%which shows that there are users who have a small number of fans but are ranked top.
To have a preliminary evaluation of these ranking results,
we compare the ranks with intrinsic qualities of the users which are 
independent of the ranking algorithm.
Specifically,
we compare the number of saved bookmarks which may represent the activity of users.
In particular,
the users {\it blackbelfjones}, {\it regine}, {\it zephoria} and {\it djakes} who appear in the top 20
of LeaderRank but not in PageRank have activity 5925, 6711, 1486 and 5082 respectively,
compared to the smaller activity 3, 377, 1516 and 242 of the users {\it thetechguy}, {\it cffcoach}, {\it samoore} and {\it kevinrose}
who appear in the top 20 of PageRank but not in LeaderRank.
This suggests that LeaderRank outperforms PageRank in identifying active users.

More detailed results are given in {\em SI}.
For instance,
the table of the top 100 users are given in Table. S1 of {\em SI}.
We have also examined the relation between scores and ranks for all the approaches,
where Zipf's laws are observed and shown in Fig.~S3 of {\em SI}. 
The overlap among the rankings obtained by LeaderRank, PageRank and the number of fans
is shown in Fig.~S4 of {\em SI}.
By comparing the relationship between the number of leaders and rank (given in Fig.~S5 of {\em SI}),
we find that PageRank tends to assign high rank to nodes with small number of leaders.
It is unfair to nodes with large number of leaders,
as users with small number of leaders are not necessarily influential
and manipulators may deliberately remove some leaders to improve their rank.
In the followings we compare, through simulations and experiments, LeaderRank,
PageRank and ranking by the number of fans.

\subsection*{Comparison with Ranking by the Number of Fans}
Ranking algorithms based on the network topology
outperform ranking by merely the number of fans.
We compare again user ranks with intrinsic qualities which are 
independent of the algorithm.
One quantity which well characterizes the user influence
is the number of times their collected bookmarks have been saved by the others.
Though the leaders are not the only sources of bookmarks,
influential users should still lead to wide spreading of their collected bookmarks.
We denote the number of bookmarks collected by user $i$ to be $B_i$
and the number of times these bookmarks are saved to be $U_i$.
A user who recommends only high quality bookmarks should have 
a large value of $U_i/B_i$.

\begin{figure}[!ht]
\leftline{\epsfig{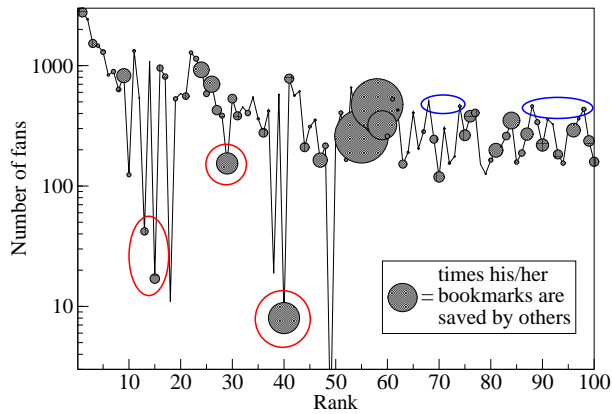} }                                                      
\caption{
The number of fans of a user in descending order of the user rank by LeaderRank.
The size of the solid circle is proportional to the value of $U_i/B_i$,
i.e. the average number of time their collected bookmarks are saved 
by others.
Users highlighted with the red circles have a small number of fans but a large value of $U_i/B_i$.
On the contrary,
users highlighted with the blue circles have a large number of fans but a small value of $U_i/B_i$.
}
\label{fig_avgSaving100}
\end{figure}

We show in \fig{fig_avgSaving100} the number of fans of a user in descending order of his/her rank by LeaderRank.
The size of the circles is proportional to the value of $U_i/B_i$.
%Upward and downward spikes correspond to users who have relatively large and small number of followers.
As we can see,
there are users who are ranked high by LeaderRank but have only a small number of fans.
Their ranks would greatly decrease if they are ranked by the number of fans.
However,
users highlighted with the red circles
have relatively large $U_i/B_i$
which shows that they are indeed high quality users.
These users are identified by LeaderRank
but not by the number of fans.
On the contrary,
there are users who have low rank but a large number of fans.
The users highlighted with the blue circles
have small $U_i/B_i$ but a large number of fans.
They are correctly ranked lower by LeaderRank.

\begin{figure}[!ht]
%\centerline{\epsfig{figure=top20.eps, width=0.5\linewidth}}
%\centerline{\epsfig{figure=top50.eps, width=0.5\linewidth}}
%\centerline{\epsfig{figure=top100.eps, width=0.5\linewidth}}
%\mbox{(a)\hspace{8.5cm}(b)}
\leftline{\epsfig{figure=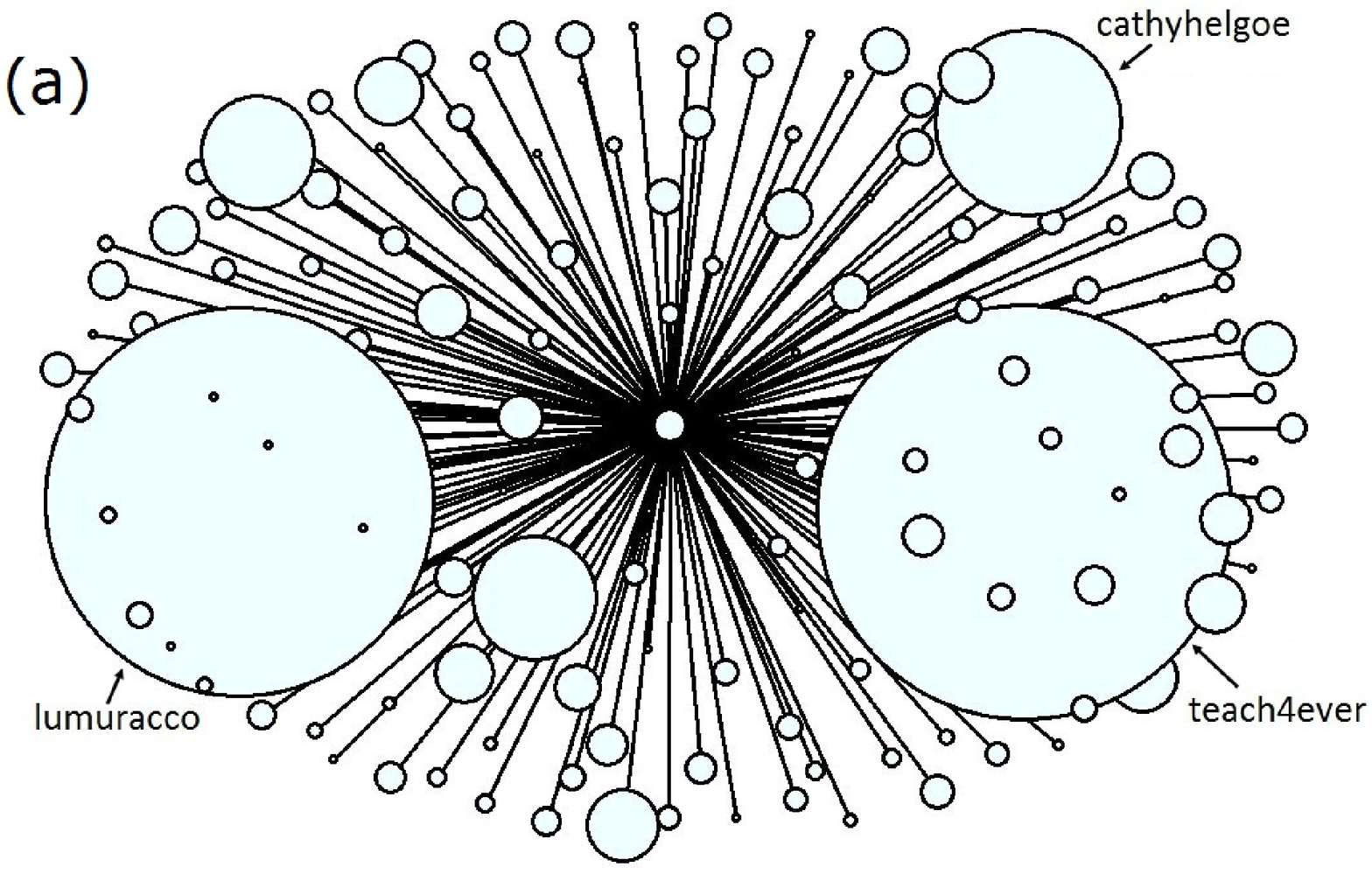, width=0.47\linewidth}                                                        
\leftline{\epsfig{figure=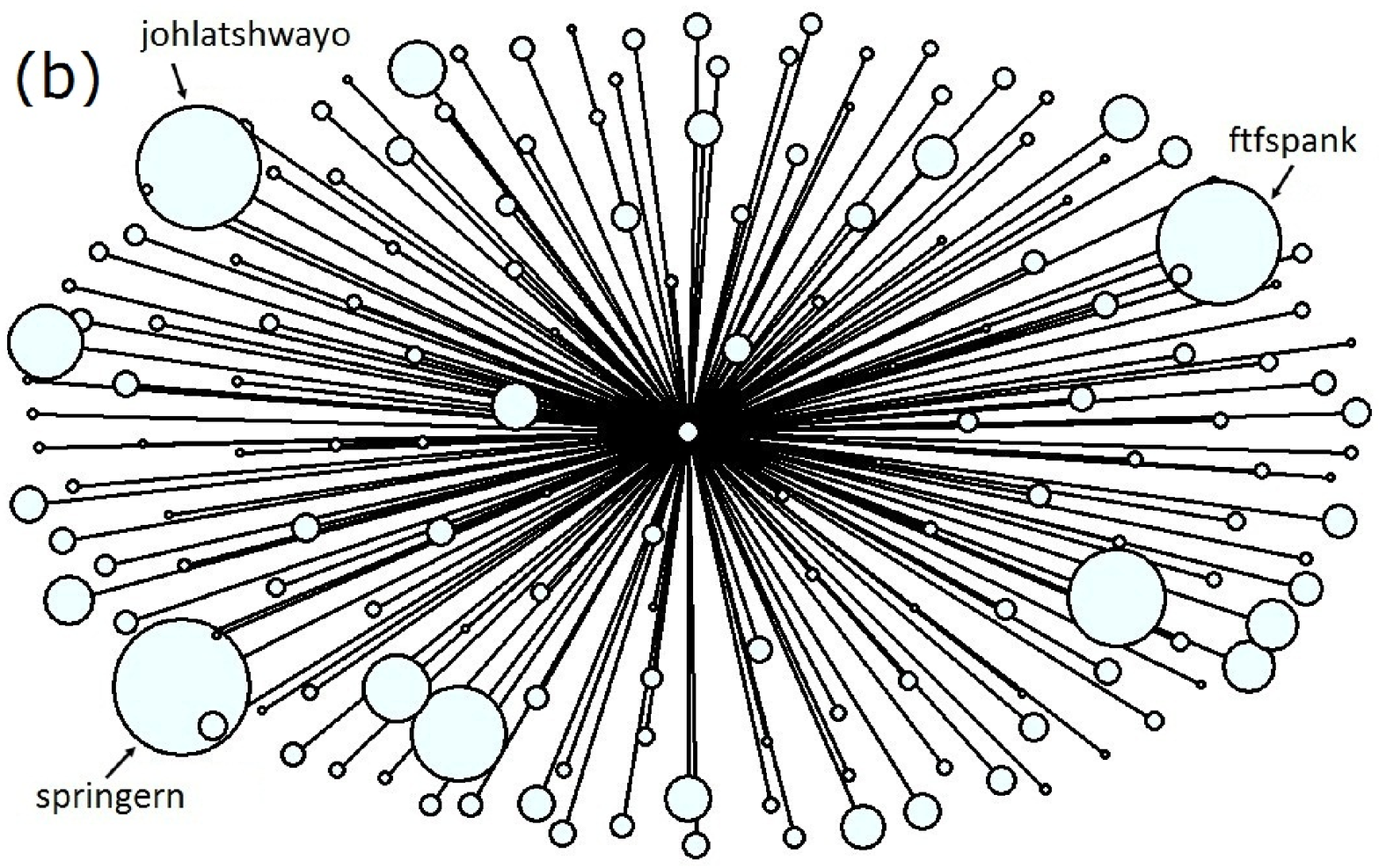, width=0.47\linewidth}}}
%\mbox{(c)\hspace{8.5cm}(d)}
\leftline{\epsfig{figure=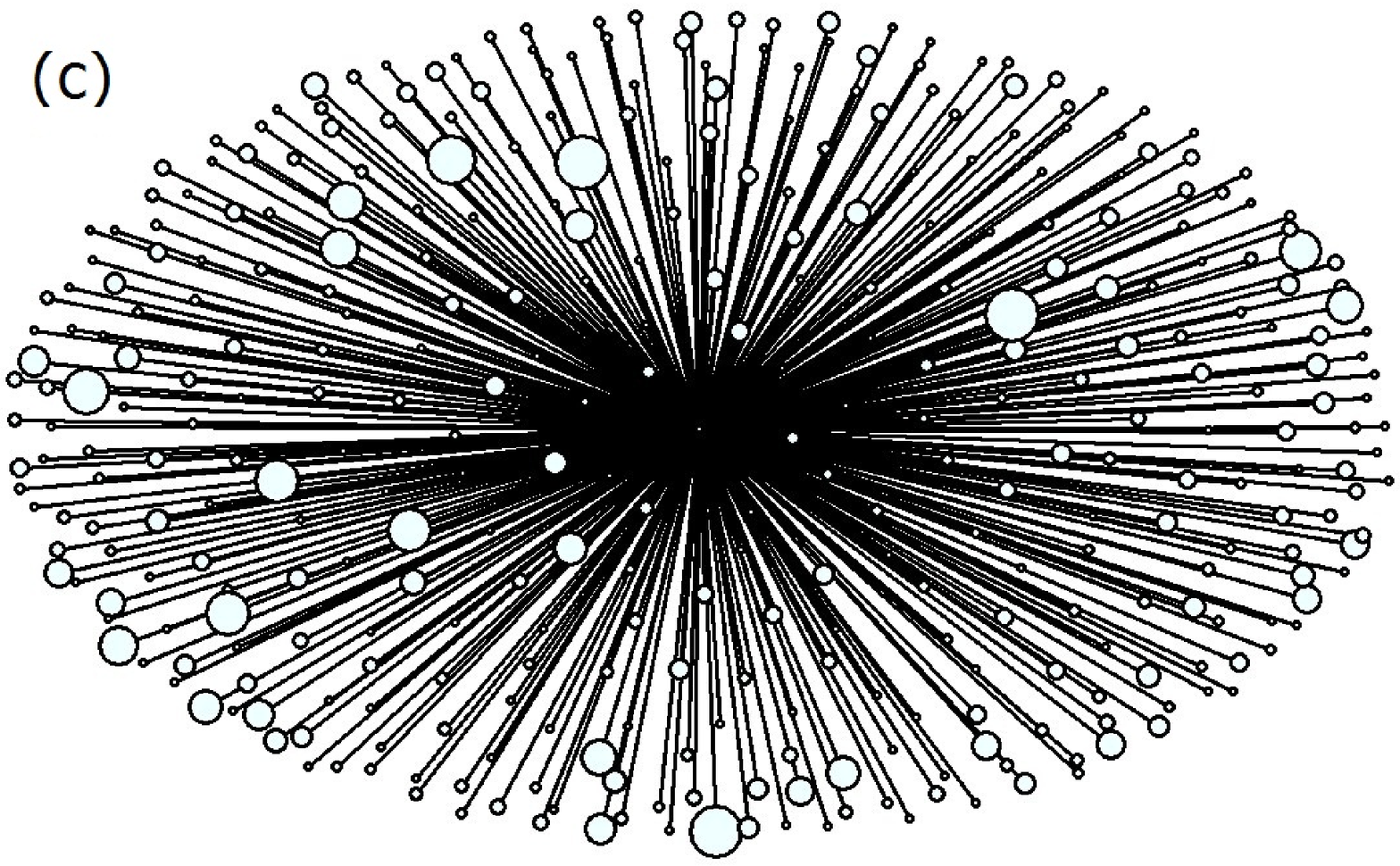, width=0.47\linewidth}                                                        
\leftline{\epsfig{figure=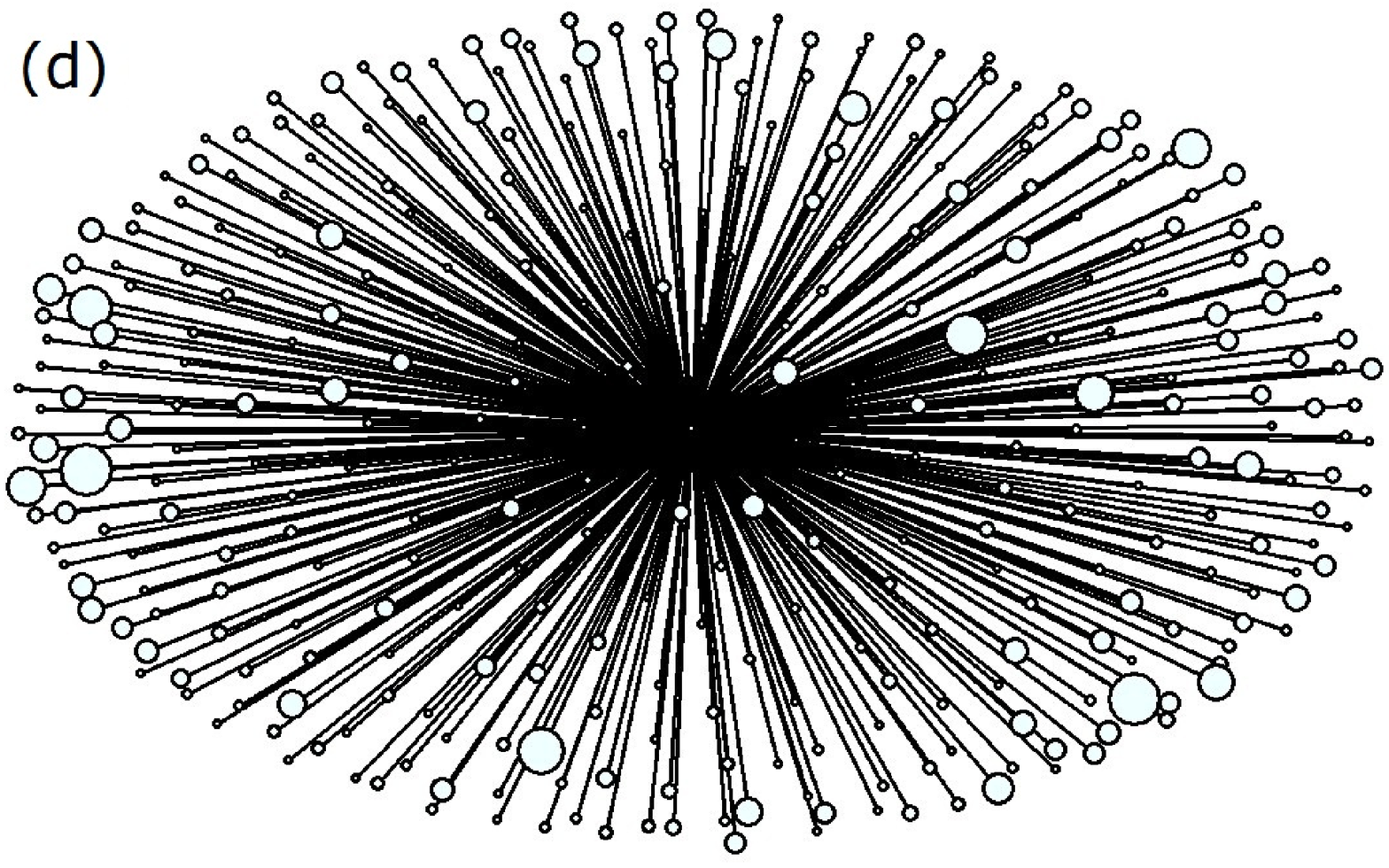, width=0.47\linewidth}}}
\caption{
Users (a) \emph{cffcoach}, (b) \emph{pedersoj}, (c) \emph{kanter} and (d) \emph{britta},
who are ranked respectively 
at $29^{\rm th}$, $47^{\rm th}$, $91^{\rm st}$ and $92^{\rm nd}$ by LeaderRank,
as surrounded by their fans.
The size of circles represents the average number of times their collected bookmarks 
are saved by others.
}
\label{fig_circle}
\end{figure}

To better understand these users,
we draw in \fig{fig_circle} particular examples of users with small number of fans but highly ranked,
and users with a large number of fans but with a relatively low rank.
As we can see in Figs. \ref{fig_circle}(a) and (b),
users \emph{cffcoach} and \emph{pedersoj} are followed by fans with large values of $U_i/B_i$,
represented by the large size of circles.
Though users \emph{kanter} and \emph{britta} have more fans,
we can see from Figs. \ref{fig_circle}(c) and (d) that they are surrounded by much smaller circles.
LeaderRank correctly gives them a lower rank,
as compared to the ranking by merely the number of fans.

Similarly,
just the leaders alone provides no absolute measure of influence,
as it is the entire upstream connection to leaders which act as the information sources
and contribute to the influence of a user.
We show in Fig. S6 of \emph{SI} that  
removing all the leaders may have a negative
effect on the social influence of a user.
All these results suggest that the leadership network is much 
more informative than simple ranking criteria such as the number of fans or leaders,
and thus algorithms which well utilize the topology can provide a better ranking.

\subsection*{Comparison with PageRank}

In addition to identifying influential users,
a good ranking algorithm for social networks should be tolerant of noisy data
and robust against manipulations.
These goals are better achieved by considering
the collective ranking based on network topology.
In the followings we compare the effectiveness and robustness between LeaderRank and PageRank,
which also utilizes topology in ranking.

\subsubsection*{Effectiveness}
How opinions spread and form in a community is an interesting question \cite{castellano09, galam02}.
To effectively spread opinion,
one has to identify influential users and create an initial social inertia.
For instance,
companies may choose to start their adverts on influential leaders
who are capable to initiate an extensive spreading through the Internet or SMS networks.
Thus a smart algorithm which ranks influential users accurately is of great commercial values.
On the other hand,
effective ranking algorithm may serve its role to identify influential users for immunization
and stop epidemic outbreak \cite{pastorSatorras02}.
As an example,
influential users who speed up junk mail spreading can be identified for targeted immunization.
Here we show that 
LeaderRank is more capable than PageRank to identify influential users who initiate a {\it quicker} and {\it wider} spreading.

\begin{figure}[!ht]
%\centerline{\epsfig{figure=top20.eps, width=0.5\linewidth}}
%\centerline{\epsfig{figure=top50.eps, width=0.5\linewidth}}
%\centerline{\epsfig{figure=top100.eps, width=0.5\linewidth}}
\leftline{\epsfig{figure=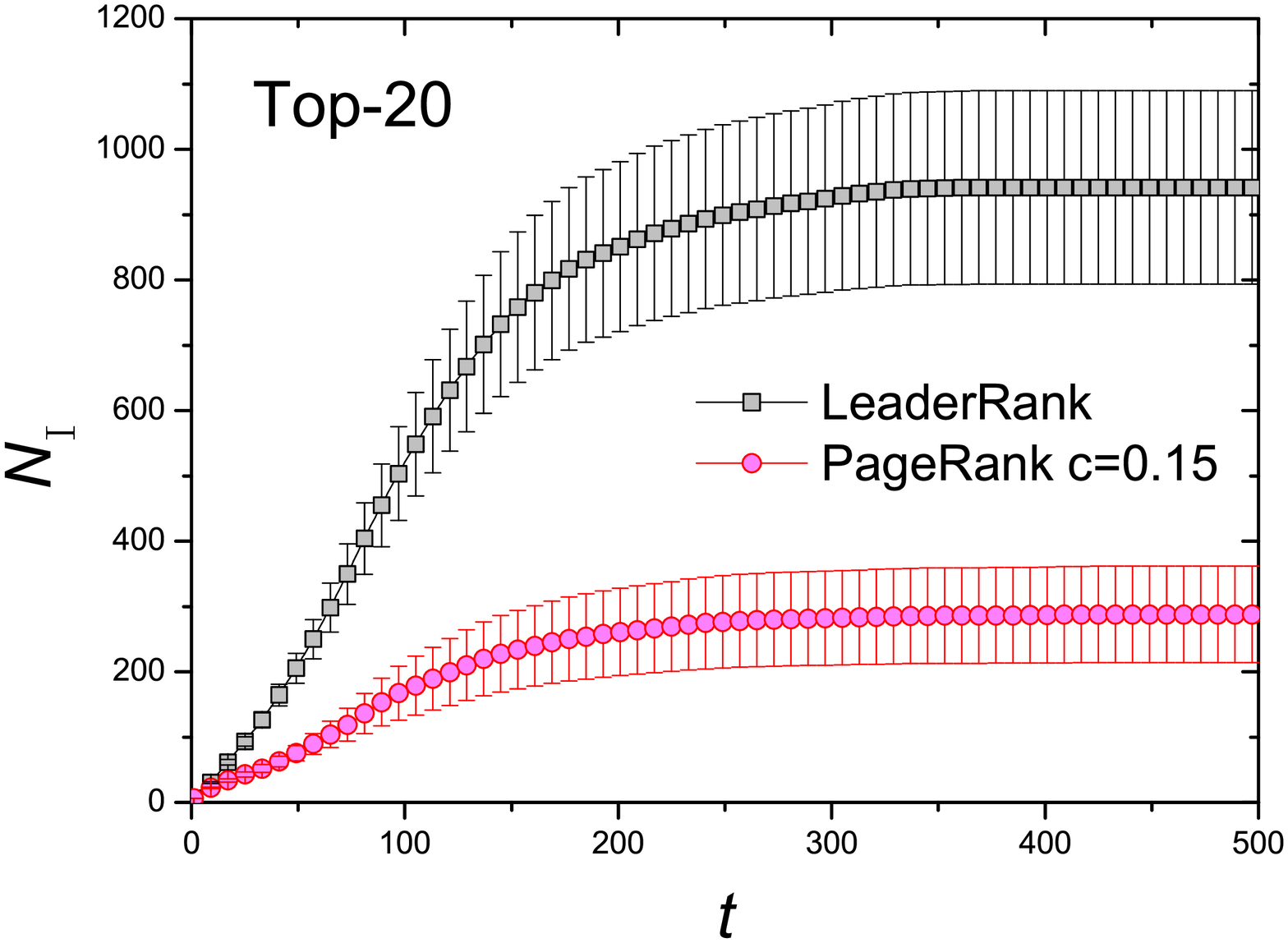, width=0.47\linewidth}                                                        
\leftline{\epsfig{figure=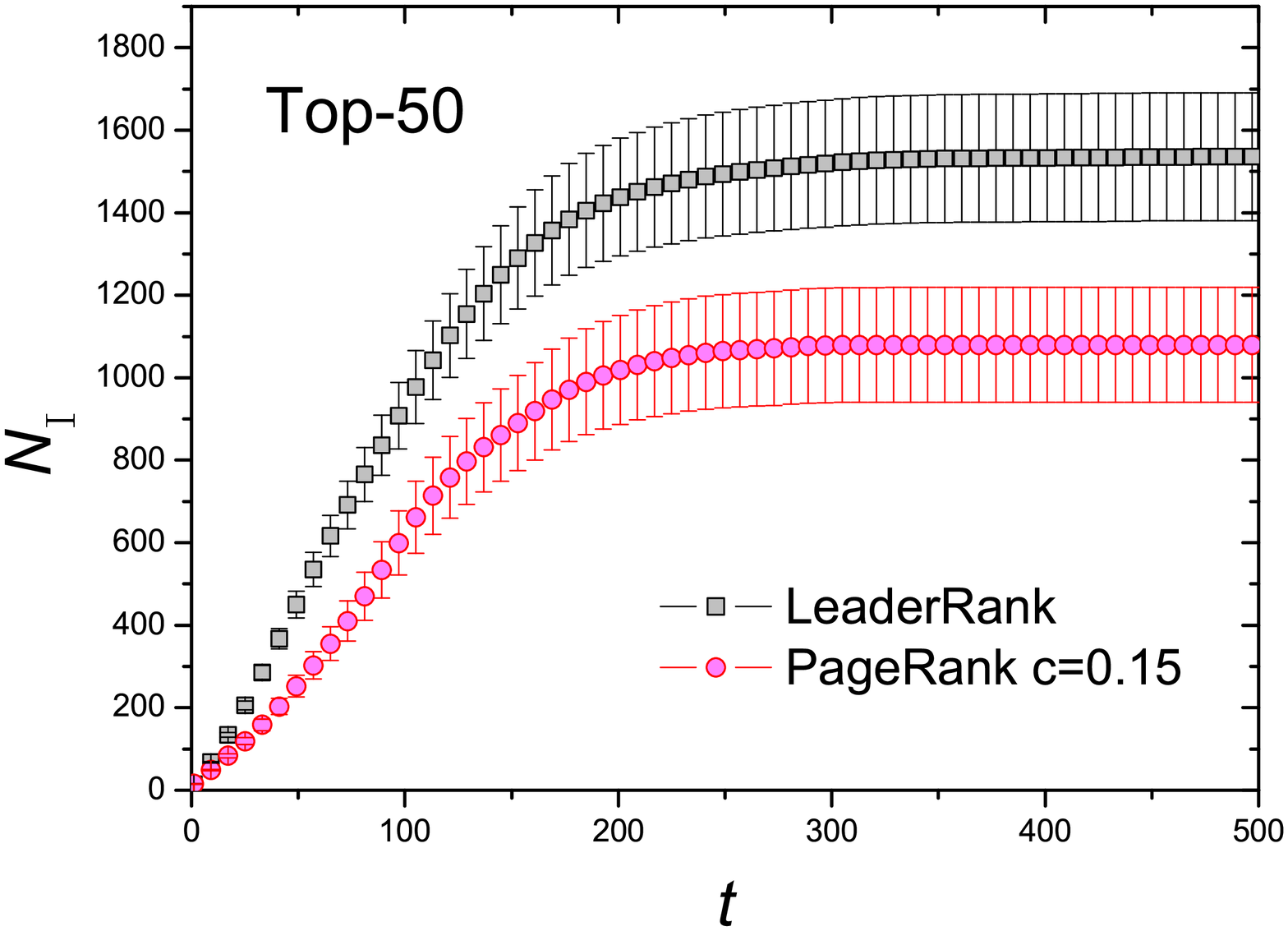, width=0.47\linewidth}}}
\leftline{\epsfig{figure=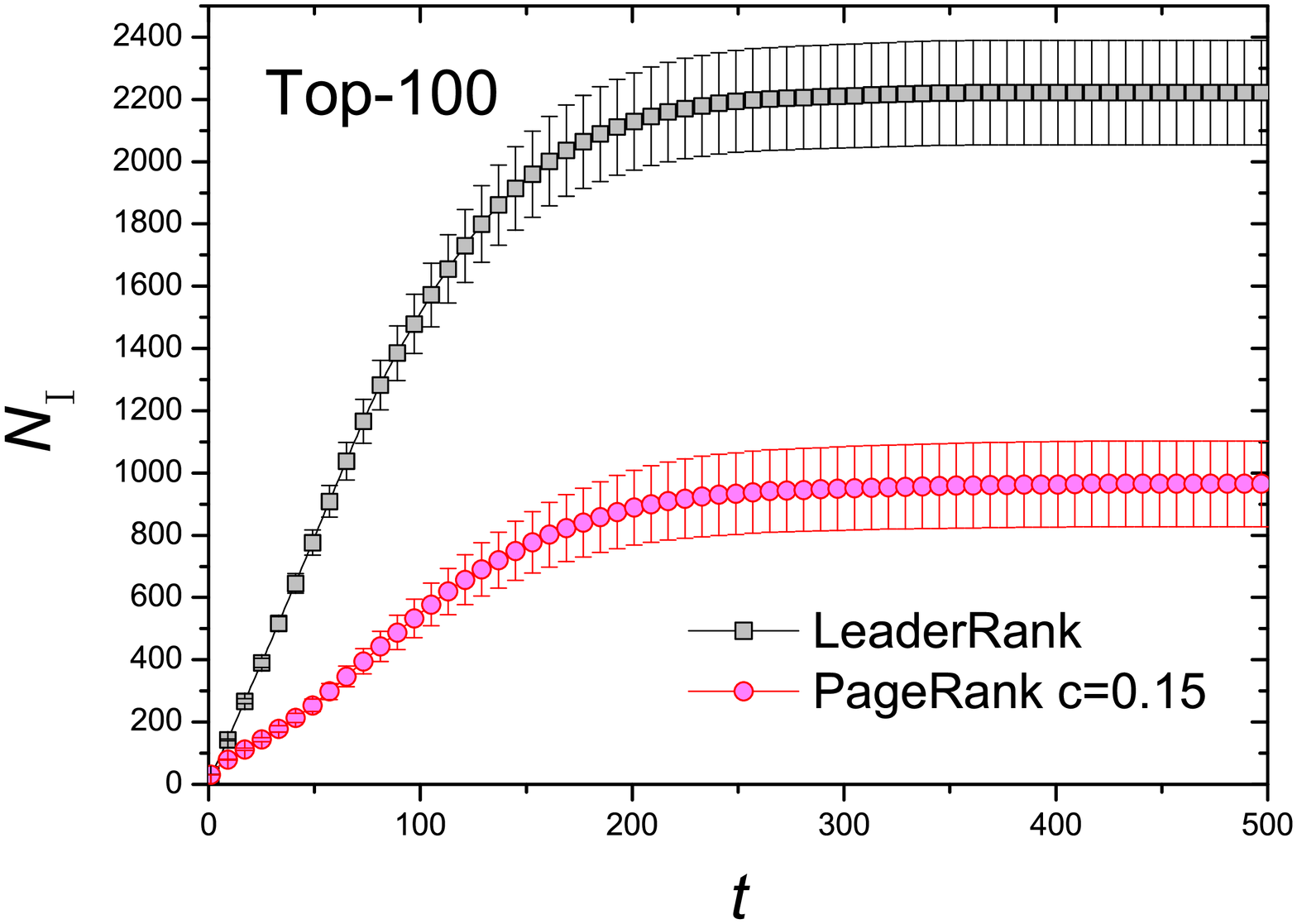, width=0.47\linewidth}                                                        
\leftline{\epsfig{figure=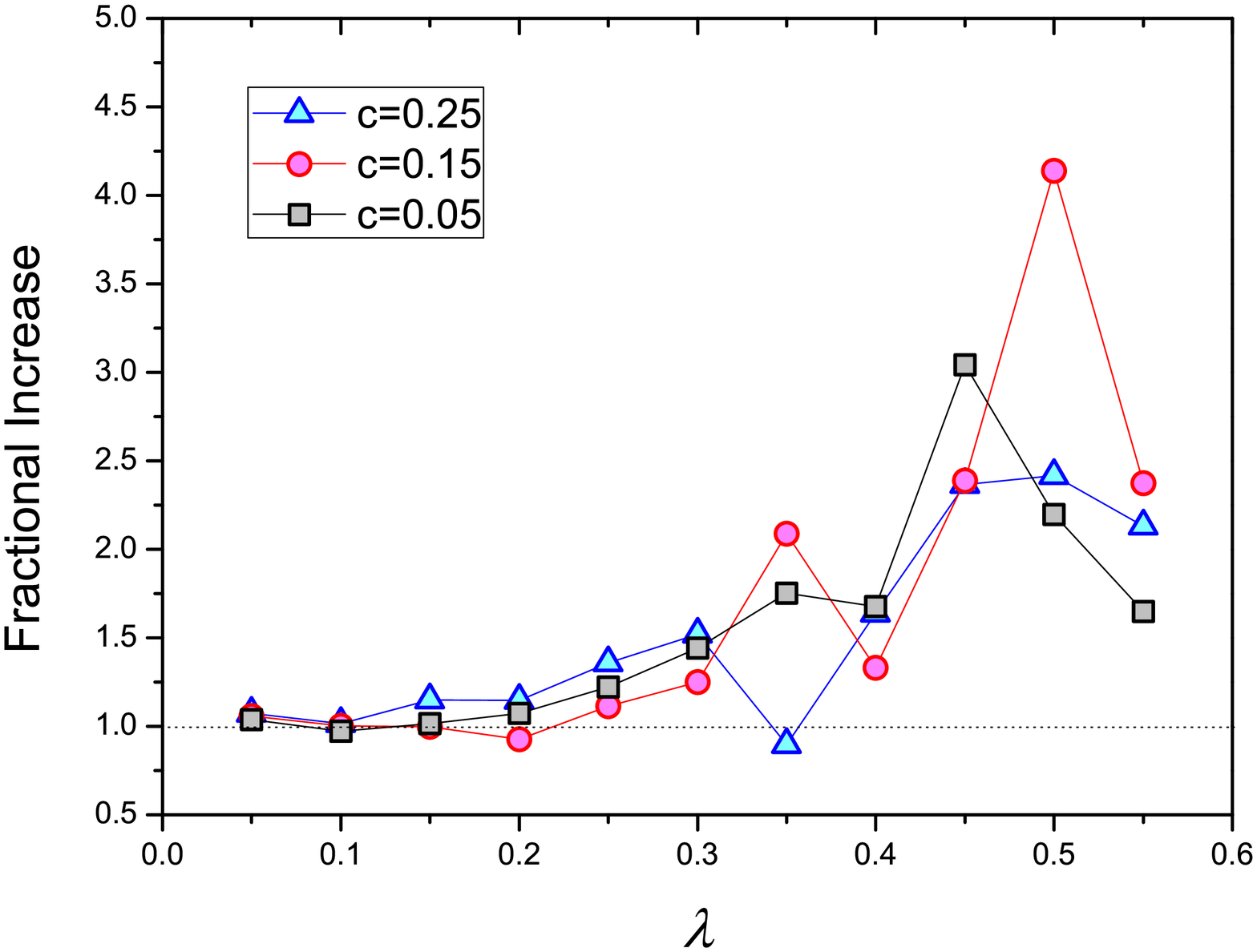, width=0.47\linewidth}}}
\caption{
The cumulative number of infected users (including recovered users), $N_{\rm I}$, as a function of time,
with initial infected to be the users either appear as 
(a) top-20,
(b) top-50,
and (c) top-100
by LeaderRank or PageRank (but not both).
As we see from \tab{tab_diffRank} in the top-20 case,
the initial infected users by LeaderRank are \emph{blackbeltjones}, \emph{regina}, \emph{zephoria} and \emph{djakes},
while that by PageRank are \emph{thetechguy}, \emph{cffcoach}, \emph{samoore} and \emph{kevinrose}.
Infection probability $\lambda=0.5$ and return probability is set to 0.15 in PageRank.
(d) As a function of $\lambda$, 
the quotient of the number of infected users in LeaderRank divided by that of PageRank,
expressed as fractional increase.
}
\label{fig_infected}
\end{figure}

Specifically,
we employ a variant of the SIR model to examine the spreading influence of the top-ranked users \cite{yang07}.
At each step,
from every infected individual,
one randomly selected fan gets infected with probability $\lambda$,
which resembles the direction of information flow.
Infected individuals recover with probability $1/\avg{k_{\rm in}}$
at each step,
where $\avg{k_{\rm in}}$ is the average in-degree of all users.
To compare the ranking effectiveness,
we set the initial infected to be the users either appear as the top 20 by LeaderRank or PageRank (but not both) in \tab{tab_diffRank},
and compare the cumulative number of infected users (which includes infected and recovered users), denoted by $N_{\rm I}$, 
as a function of time.
The initial infected users by the two algorithms are given in the caption of \fig{fig_infected}.
This experiment resembles an opinion spreading initiated from the top users and observe how the opinion propagates. 
Figure \ref{fig_infected}(a) shows that 
infecting the top users from LeaderRank 
results in a faster growth and a higher saturated number of infected,
indicating a {\it quicker} and {\it wider} spreading.
To further confirm the effectiveness of LeaderRank,
we also conduct experiments for the top 50 and top 100 ranked users either from LeaderRank or PageRank
and obtain similar results which are shown in Figs. \ref{fig_infected}(b) and (c),
respectively.

We show in \fig{fig_infected}(d) 
the quotient of the total infected in LeaderRank divided by that of PageRank,
with different infection probability $\lambda$.
LeaderRank outperforms PageRank of various return probability
and for a broad indicated range of $\lambda$.
This reveals again a drawback of PageRank
as the optimal return probability has to be found by extensive parameter tests.
The results imply that spreading from both LeaderRank and PageRank users 
is limited when $\lambda$ is small,
but LeaderRank leads to a much wider opinion spreading when $\lambda$ is large.
For a virus outbreak,
if intensive immunizations are implemented on the top ranked LeaderRank users,
the final outbreak would be less extensive.
All the above results show that LeaderRank is more effective than PageRank
in identifying highly influential users,
and is thus a better candidate for opinion spreading and to prevent a virus outbreak.

\subsubsection*{Tolerance of Noisy Data}
Tolerance of ranking against spurious and missing links,
i.e. false positive and false negative connections,
is crucial when network structure is subject to noisy observations \cite{guimera09}.
Social network data may be unreliable,
especially when users are required to explicitly indicate relationship with others \cite{marsden90}. 
It is like,
to state whether neighbors are friends if they just greet each other when they meet.
The same happens for networks other than social networks
but with a rather different cause.
For example,
protein connections obtained from biological experiments 
often include numerous false positives and false negatives \cite{legrain01}.
Other than ambiguous personal relationship,
it is also costly and technically difficult to explore social networks comprehensively.
Efforts have thus been made to predict the missing connections \cite{lu10} and on such noisy networks,
we should develop ranking algorithms which are tolerant of spurious and missing links.

\begin{figure}[!ht]
\leftline{\epsfig{figure=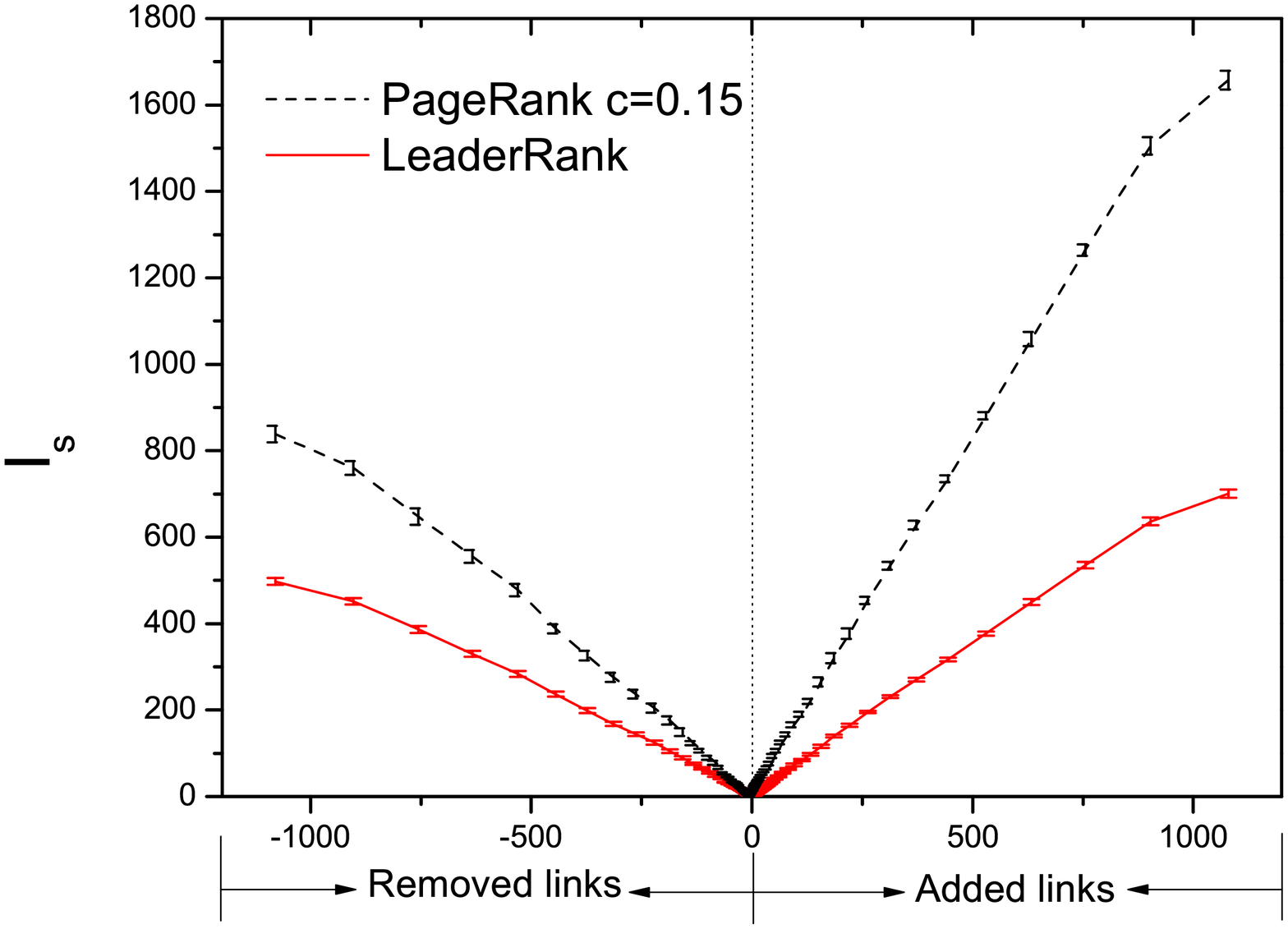, width=0.47\linewidth}                                                        
\leftline{\epsfig{figure=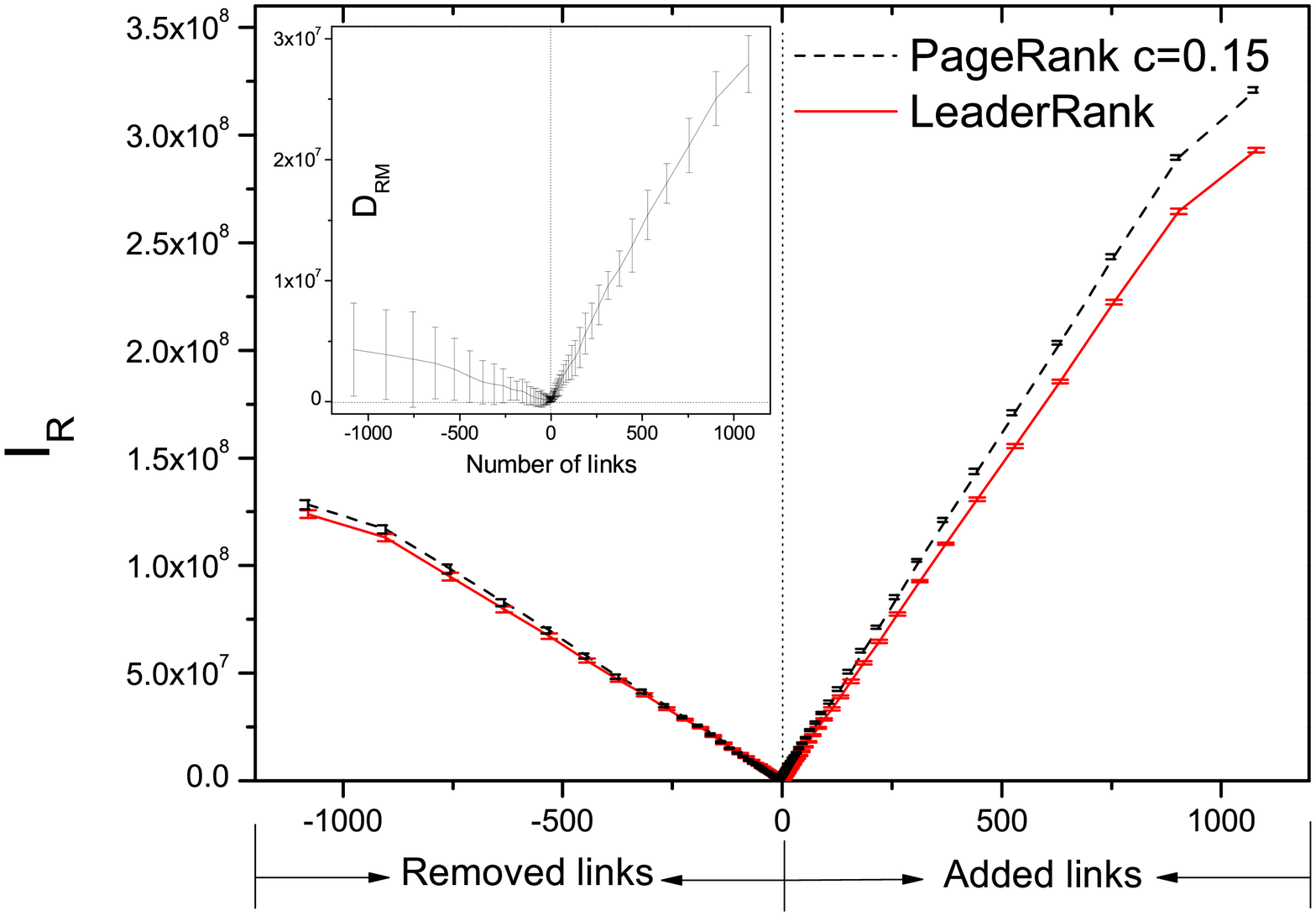, width=0.47\linewidth}}}
\caption{
The impact on (a) scores and (b) ranking as a function
of number of links added and removed.
Inset: (b) the difference in ranking mobility between LeaderRank and PageRank.
}
\label{fig_rankStable}
\end{figure}

To examine the tolerance of LeaderRank and PageRank against noisy data,
we measure the change in scores and rankings when links are added
or removed randomly.
These links correspond to the spurious or missing relationship among leaders and fans.  
The scores obtained from the modified graph are compared to
those from the original graph,
by measuring the impact $I_S$ on score,
as given by
\begin{eqnarray}
	I_S = \sum_{i=1}^{N}|S'_i-S_i|,
\end{eqnarray}
where $S_i$ and $S'_i$ correspond to the scores obtained 
respectively from the original and modified graph.
We measure $I_S$ for both LeaderRank and PageRank
subject to the same modifications.
As shown in \fig{fig_rankStable}(a),
$I_S$ increases with the number of links added or 
removed.
Remarkably,
much smaller values of $I_S$ are obtained from LeaderRank
when compared to PageRank,
regardless of the addition or removal of links.
In a word,
LeaderRank is more tolerant than PageRank against noisy topology,
and thus has a high potential
in applications on noisy social networks or protein-protein networks \cite{chen09}.

Since a small change in scores in LeaderRank may not directly correspond to a small change in ranking,
we define a similar measure to examine the impact $I_R$ on ranking,
given by
\begin{eqnarray}
	I_R = \sum_{i=1}^{N}|R'_i-R_i|.
\end{eqnarray}
As shown in \fig{fig_rankStable}(b),
a smaller difference between $I_R$ of LeaderRank and PageRank 
is observed as compared to $I_S$.
Nevertheless $I_R$ of LeaderRank is smaller,
as shown by $D=I^{\rm Page}_R-I^{\rm Leader}_R>0$ in the inset.
Once again,
these observations in $I_R$ suggest that LeaderRank is more tolerant of
topology randomness and hence a better candidate for ranking in noisy networks.

\subsubsection*{Robustness against Spammers}
Malicious activities are common in social networks,
in particular when users manipulate to gain skewed reputation \cite{masum04}.
One example of manipulation is called  {\it Sybil Attack} \cite{douceur02},
in which spammers deliberately create fake entities to obtain disproportionately high rank.
The problems become intolerable if this manipulation causes recommendation
of bad commodities or biased opinion in social networks.
In WWW,
there are also stories of companies manipulating Google search engine
to obtain higher ranks in search results \cite{levine06}.
To cope with this loophole,
we show that LeaderRank is more robust than PageRank against this type of attacks.

\begin{figure}[!ht]
%\centerline{\epsfig{figure=spam_fr.eps, width=0.5\linewidth}}
%\centerline{\epsfig{figure=spam_pr.eps, width=0.5\linewidth}}
\leftline{\epsfig{figure=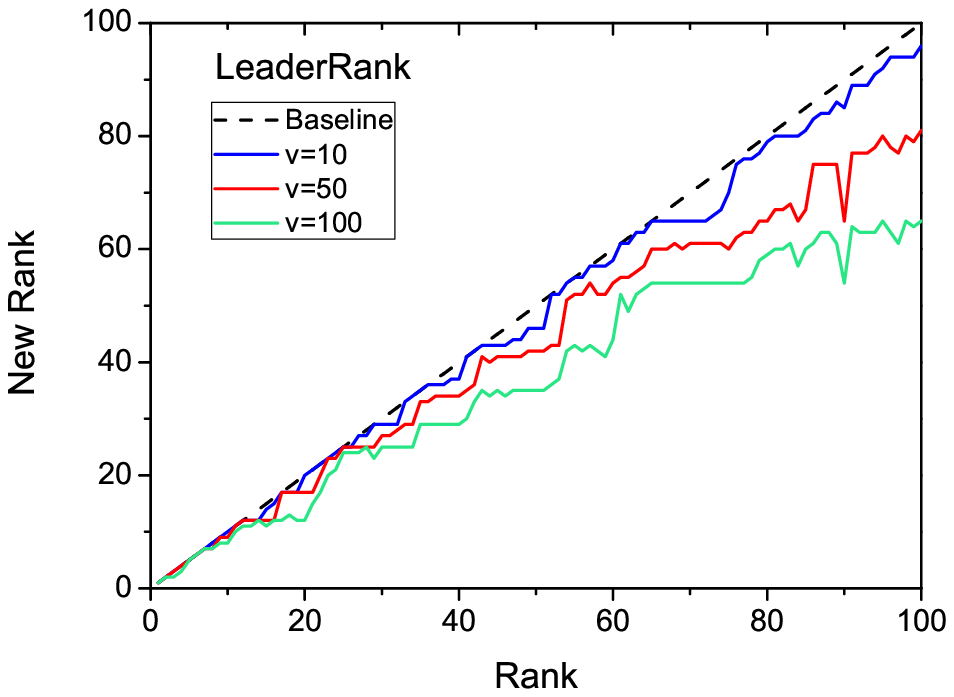, width=0.47\linewidth}                                                        
\leftline{\epsfig{figure=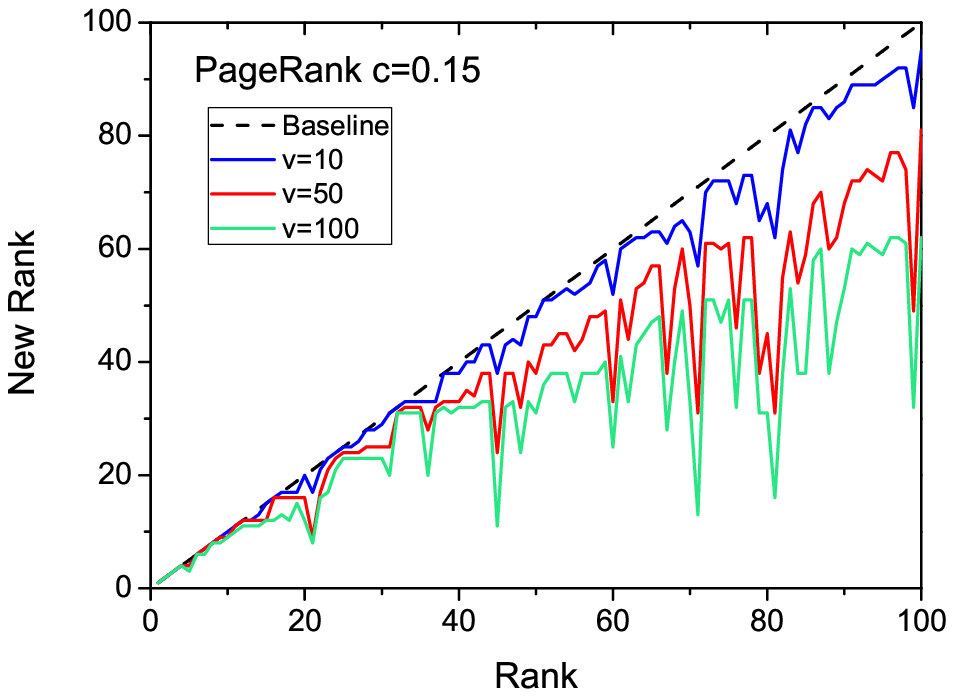, width=0.47\linewidth}}}
\caption{
The manipulated rank as obtained by (a) LeaderRank and (b) PageRank,
after the addition of $v$ fake fans, 
with $v=10, 50, 100$.
}
\label{fig_rank}
\end{figure}

Specifically,
we simulate the situation where a user creates $v$ fake fans,
and compare the ranking robustness in LeaderRank and PageRank.
The horizontal axis of Figs. \ref{fig_rank}(a) and (b) shows respectively 
for LeaderRank and PageRank the original rank of a user,
and the vertical axis shows his/her
manipulated rank after the addition of $v$ fake fans.
Vertical downward shift from the dashed diagonal corresponds to 
the increase in rankings,
and thus a successful manipulation.
As we can see,
LeaderRank is more robust against spammers as the change of rankings
is much smaller than that by PageRank.
These results show that LeaderRank is a better candidate for robust rankings 
against manipulations.

\subsection*{Experiment}

To let readers better understand social influences
as quantified by LeaderRank,
we established a webpage 
\url{http://rank.sesamr.com} which uses LeaderRank to rank users in delicious.com.
By providing their username,
delicious users can easily obtain their rank 
and other information including the influence of leaders and fans.
Users can also examine the change of their influence
when they have new leaders and fans.
For instance,
the user \emph{babyann519} had a low rank of 607512
before six other users found her important bookmarks and added her as a leader.
She now has a rank of 99440, 
a much higher rank which shows the increase in her influence.

\section*{Discussion}

After going through the above details,
we may conclude that identifying influential users is not a simple task.
It is not merely answering who is the best,
but as well to consider 
the influences and consequences brought by a ranking algorithm.
These consequences are of particular importance for social networks,
which are fundamentally different from networks of webpages.
For instance,
the ranking should be robust against noisy data
and smart manipulations.
This leads us to answer a much broader question by devising a robust and generic algorithm,
than merely identifying the leaders.

We suggest that LeaderRank may serve as a prototype
of ranking algorithms applicable to rank users in social networks.
As personal relationships are quickly evolving,
the adaptive and parameter-free nature of LeaderRank eliminates
the need of frequent calibration.
In addition,
this simple algorithm outperforms PageRank in several important aspects.
In this paper,
we see that LeaderRank identifies users who lead to quick and extensive 
spreading of opinions.
This is important for online applications
which feature information spreading.
On the other hand,
LeaderRank is tolerant of spurious and missing links,
which benefits applications with noisy data,
especially personal relationship.
To deal with ranking loopholes,
LeaderRank is robust against manipulations.
These results make LeaderRank a good candidate
for ranking users as well as other ranking tasks.

Though LeaderRank is already an effective algorithm,
extensions may lead to further improvement.
For instance,
the role of the ground node would be more prominent if weights are
set on the in- and out-links to each node,
according to its significance or other criteria.
It can also be generalized to applications
ranging from blog plagiarizer identification \cite{gayo-avello10},
to stopping species lost in ecosystem \cite{allesina08}.
These simple modifications may lead to substanial improvements in performance.

Identifying influential users in social networks 
is still a task on which we may overlook.
As accompanied by the expanding popularity of online communities,
leader identification may reinforce
their development.
This further facilitates collective search through online communities 
and may one day complement the current search paradigm.
For sure in the near future, 
technological advance will
provide more information to quantify user influence,
but at the same time will scale up the network size and make ranking tasks
more challenging.
LeaderRank suggested here may serve as a potential candidate to face this challenge
and well utilize the power of social influences.

% Do NOT remove this, even if you are not including acknowledgments
\section*{Acknowledgments}
This work is supported by 
the Shanghai leading discipline project (under grant S30501),
QLectives projects (EU FET-Open Grants 213360 and 231200),
National Natural Science Foundation of China under Grant Nos. 10635040 and 90924011
and the Swiss National Science Foundation (200020-132253).
We thank Zi-Ke Zhang for the data preparation, Hu Xia for the data analysis, You-Gui Wang for introducing the concept of mobility, 
Dong Wei and Hao Liu for implementing the website,
Mat\'u\v s Medo and Joseph Wakeling for fruitful discussions.

%\section*{References}
% The bibtex filename
%\bibliography{template}

\begin{thebibliography}{}

\bibitem{lampe06}
Lampe C, Ellison N, Steinfield C (2006)
A Face(book) in the crowd: social searching vs. social browsing.
Proceedings of the 20th anniversary conference on computer supported cooperative work.
pp 167-170.

\bibitem{vieira07}
Vieira MV, Fonseca BM, Damazio R, Golgher PB, de Castro Reis D, Ribeiro-Neto B (2007)
Efficient search ranking in social networks.
Proceedings of the 16th ACM conference on information and knowledge management.
pp 537-572.

\bibitem{easleyBook}
Easley D, Kleinberg J (2010)
Networks, Crowds and Markets.
Cambridge University Press, New York.

\bibitem{kleinberg99}
Kleinberg J (1999) 
Authoritative sources in a hyperlinked environment. 
Journal of ACM 46:604-632.

\bibitem{park05}
Park J and Newman MEJ (2005)
A network-based ranking system for US college football.
J Stat Mech P10014.

\bibitem{radicchi09}
Radicchi F, Fortunato S, Benjamin M, Vespignani A (2009)
Diffusion of scientific credits and the ranking of scientists.
Phys Rev E 80:056103.

\bibitem{chen07}
Chen P, Xie H, Maslov S, Redner S (2007)
Finding scientific gems with Google.
J Inform 1:8-15.

\bibitem{page99}
Page L, Brin S, Motwani R, Winograd T (1999) 
The PageRank citation ranking: Bringing order to the web. 
Technical Report Stanford InfoLab 1999-66.

\bibitem{brin98}
Brin S, Page L (1998)
The anatomy of a large-Scale hypertextual web search engine.
Comput Networks ISDN 30:107-117.

\bibitem{masum04}
Masum H, Zhang YC (2004)
Manifesto for the reputation society.
First Monday 9:7-5.

\bibitem{watts1998}
Watts DJ, Strogatz SH (1998) 
Collective dynamics of `small world' networks.
Nature 393:440-442.

\bibitem{newman2002}
Newman MEJ (2002)
Assortative mixing in networks.
Phys Rev Lett 89:208701.

\bibitem{castellano09}
Castellano C, Fortunato S, Loreto V (2009)
Statistical physics of social dynamics.
Rev Mod Phys 81:591-646.

\bibitem{galam02}
Galam S (2002)
Minority opinion spreading in random geometry.
Eur Phys J B 25:403-406.

\bibitem{pastorSatorras02}
Pastor-Satorras R, Vespignani A (2002)
Immunization of complex network.
Phys Rev E 65:036104.

\bibitem{yang07}
Yang R, Wang BH, Ren J, Bai WJ, Shi ZW, Wang WX, Zhou T (2007)
Epidemic spreading on heterogenous networks with identical infectivity.
Phys Lett A 364:189-193.

\bibitem{guimera09}
Guimer\'a R, Sales-Pardo M (2009)
Missing and spurious interactions and the reconstruction of complex networks.
PNAS 106:22073-22078. 

\bibitem{marsden90}
Marsden PV (1990)
Network data and measurement.
Annual Review of Sociology 16:435-463.

\bibitem{legrain01}
Legrain P, Wojcik J, Gauthier JM (2001)
Protein¡Vprotein interaction maps: a lead towards cellular functions.
Trends in Genetics 17:346-352.

\bibitem{lu10}
L\"u L, Zhou T (2011)
Link prediction in complex network: A survey.
Physica A 390:1150-1170.

\bibitem{chen09}
Chen J, Aronow BJ, Jegga AG (2009)
Disease candidate gene identification and prioritization using protein interaction networks,
BMC Bioinformatics 10:73. 

\bibitem{douceur02}
Douceur JR (2002)
The Sybil Attack.
Proceedings of the First International Workshop on Peer-to-Peer Systems.

\bibitem{levine06}
Levine BN, Shields C, Margolin NB (2006)
A survey of solutions to the sybil attack. 
Technical Report of Univ of Massachussets Amherst 2006-052.


\bibitem{gayo-avello10}
Gayo-Avello D (2010)
Nepotistic relationships in Twitter and their impact on rank prestige algorithms.
arxiv.org 1004.0816.

\bibitem{allesina08}
Allesina S and Pascual M (2009)
Googling food webs: Can an eigenvector measure species' importance for coextinctions?
PLoS Comput Bio 5:e1000494.


\end{thebibliography}

\newpage

\section*{Figure Legends}
%\begin{figure}[!ht]
%\begin{center}
%%\includegraphics[width=4in]{figure_name.2.eps}
%\end{center}
%\caption{
%{\bf Bold the first sentence.}  Rest of figure 2  caption.  Caption 
%should be left justified, as specified by the options to the caption 
%package.
%}
%\label{Figure_label}
%\end{figure}

%%%%%%%%%%%%%%%%%%%%%%%%%%%%%%%%
%%%%%%%%%%%%%%%%%%%%%%%%%%%

%%%%%%%%% Figure 5 %%%%%%%%%%%%%%%%%
%%%%%%%%%%%%%%%%%%%%%%%%%%%%%%%%%%%%

%%%%%%%%% Figure 6 %%%%%%%%%%%%%%%%%
%%%%%%%%%%%%%%%%%%%%%%%%%%%%%%%%%%%%

%%%%%%%%% Figure 6 %%%%%%%%%%%%%%%%%
%%%%%%%%%%%%%%%%%%%%%%%%%%%%%%%%%%%%

%%%%%%%%% Figure 5 %%%%%%%%%%%%%%%%%
%%%%%%%%%%%%%%%%%%%%%%%%%%%%%%%%%%%%

%%%%%%%%% Figure 6 %%%%%%%%%%%%%%%%%
%%%%%%%%%%%%%%%%%%%%%%%%%%%%%%%%%%%%

%\section*{Tables}
%\begin{table}[!ht]
%\caption{
%\bf{Table title}}
%\begin{tabular}{|c|c|c|}
%table information
%\end{tabular}
%\begin{flushleft}Table caption
%\end{flushleft}
%\label{tab:label}
% \end{table}

%%%%%%%%%%%%% Table top-20 users %%%%%%%%%%%%%%
\begin{table}[!ht]
\caption{Top 20 users ranked by the three approaches.} 
\label{tab_diffRank}
%\rowcolors{4}{gray}{}
\begin{tabular}{|c|c|c|c|}
\hline
%\multirow{2}{*}{User ID}
User ID
& \multicolumn{3}{|c|}{Ranking} \\
\cline{2-4}
& LeaderRank & PageRank & 
By the number of fans \\
\hline \hline
adobe & 1 & 1 & 1  \\
\rowcolor{gray!20}
twit & 2 & 2 & 2 \\
wfryer & 3 & 6 & 3 \\
\rowcolor{gray!20}
willrich &4 &7 &4 \\
joshua & 5 & 8 & 6 \\
\rowcolor{gray!20}
cshirky & 6 & 12 & 13 \\
hrheingold &7 & 15 & 12 \\
\rowcolor{gray!20}
ewan.mcintosh & 8 & 14 & 19 \\
dwarlick & 9 & 19 & 14 \\
\rowcolor{gray!20}
twitarmy & 10 & 3 & \\
merlinmann & 11 & 16 & 5 \\
\rowcolor{gray!20}
blackbeltjones & 12 & & \\
jdehaan & 13 & 9 & \\
\rowcolor{gray!20}
regine & 14 & & 9 \\
lseymour & 15 & 10 & \\
\rowcolor{gray!20}
jonhicks & 16 & 17 & 10 \\
zephoria & 17 & & 15 \\
\rowcolor{gray!20}
isola & 18 & 11 & \\
djakes & 19 & & \\
\rowcolor{gray!20}
secondlife & 20 & 13 & \\
thetechguy & & 4 & \\
\rowcolor{gray!20}
cffcoach & & 5 & \\
samoore & &18 & \\
\rowcolor{gray!20}
kevinrose & & 20 & 11 \\
steverubel & & & 7 \\
\rowcolor{gray!20}
jgwalls & & & 8 \\
ambermac & & & 16 \\
\rowcolor{gray!20}
jgates513 & & & 17 \\
ramitsethi & & & 18 \\
\rowcolor{gray!20}
cory$\_$arcangel  & & & 20 \\
\hline
\end{tabular}
\end{table}
%%%%%%%%%%%%%%%%%%%%%%%%%%%%%%%%%%%%%%%%%%%%%

\clearpage
\setcounter{figure}{0}
\setcounter{table}{0}
\setcounter{equation}{0}
\renewcommand{\req}[1]{Eq.~(S\ref{#1})}
\renewcommand{\fig}[1]{Fig.~S\ref{#1}}
\renewcommand{\tab}[1]{Table S\ref{#1}}
\renewcommand{\figurename}{Fig.~S\hspace{-0.1cm}}
\renewcommand{\tablename}{Table~S\hspace{-0.1cm}}
\makeatletter
\def\@eqnnum{{\normalfont \normalcolor (S\theequation)}}
\makeatother

%%%%%%%%%%%%%%%%%%%%%%%%%%%%%%

%\title{Supporting Information}
\centerline{\LARGE Supporting Information}

\section{Primitivity and Convergence}
We first show that the stochastic matrix $P$ is primitive by showing $P^6$ is positive,
i.e. the elements in $P^6$ are all greater than zero.
It is equivalent to show that any pair of nodes are connected in exactly 6 steps (6 hops).
For nodes with at least one link,
ground node guarantees the co-existence of loops of size 2 and 3.
Starting at any node with 2 loops of size 2 and a path through the ground node,
we can reach any other node (excluding the ground node but including itself) in exactly 6 steps.
To reach the ground node in exactly 6 steps,
we make use of one loop of size 3 and one loop of size 2 before hopping to the ground node.
The same is true to reach the other nodes from the ground node.

As $P$ is a right stochastic matrix,
the transpose $P^{T}$ would be the usual transition matrix by conventional matrix multiplication,
such that $\vec{s}(t_c)=P^T\vec{s}(t_c)$.
We then show that 1 is an eigenvalue of $P$,
and thus of $P^T$.
The matrix $P$, 
which is row-normalized,
has obviously an eigenvalue 1 with eigenvector filled with all equal entries,
and thus 1 is an eigenvalue of $P$.
To show the uniqueness of eigenvector associated with eigenvalue 1,
we assume that there exists another eigenvector $\vec{v}$ for eigenvalue 1 with heterogeneous entries.
Let $v_j$ to be the entry of this eigenvector with $|v_j|>|v_i|$ for all $i$.
We then choose the eigenvector such that $v_j$ is positive.
As $P$ is primitive,
we consider a matrix $P^m$ where all entries are positive.
The assumption of eigenvector with heterogeneous entries leads to the following contradiction
\begin{eqnarray}
	\vec{v} = P^m\cdot\vec{v} \Rightarrow v_j=\sum_{i}p'_{ij} v_i<\sum_{i}p'_{ij} v_j=v_j,
\end{eqnarray}
where $p'$ denotes the elements of $P^m$.
The contradiction implies that for $P^m$,
and hence $P$,
the eigenvector with heterogeneous entries does not exist for eigenvalue 1,
and thus $P^T$ has an unique eigenvector associated with eigenvalue 1,
i.e. a unique steady state.

\section{Differences between LeaderRank and PageRank}

\begin{figure}
\leftline{\epsfig{figure=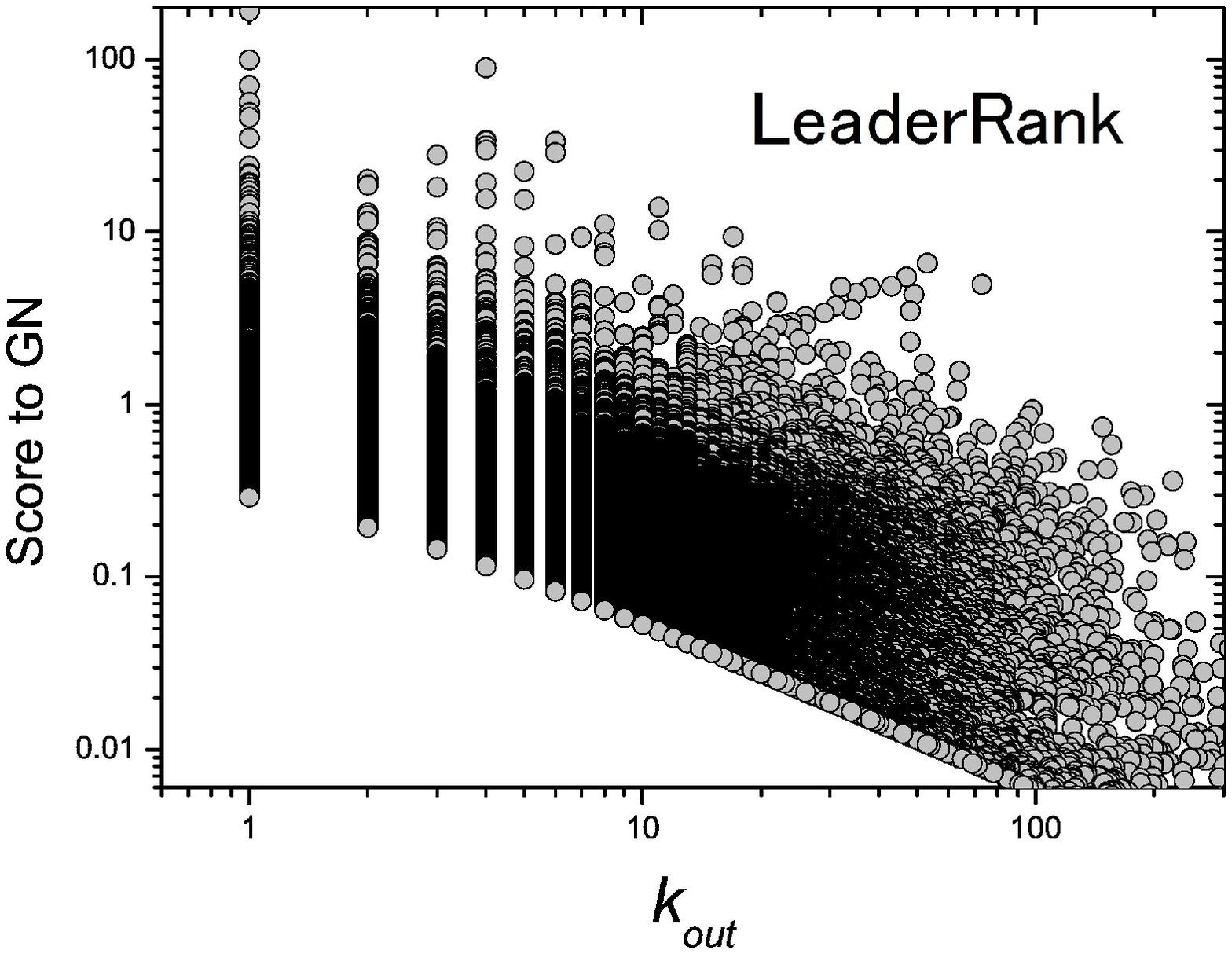, width=0.47\linewidth}                                                        
\leftline{\epsfig{figure=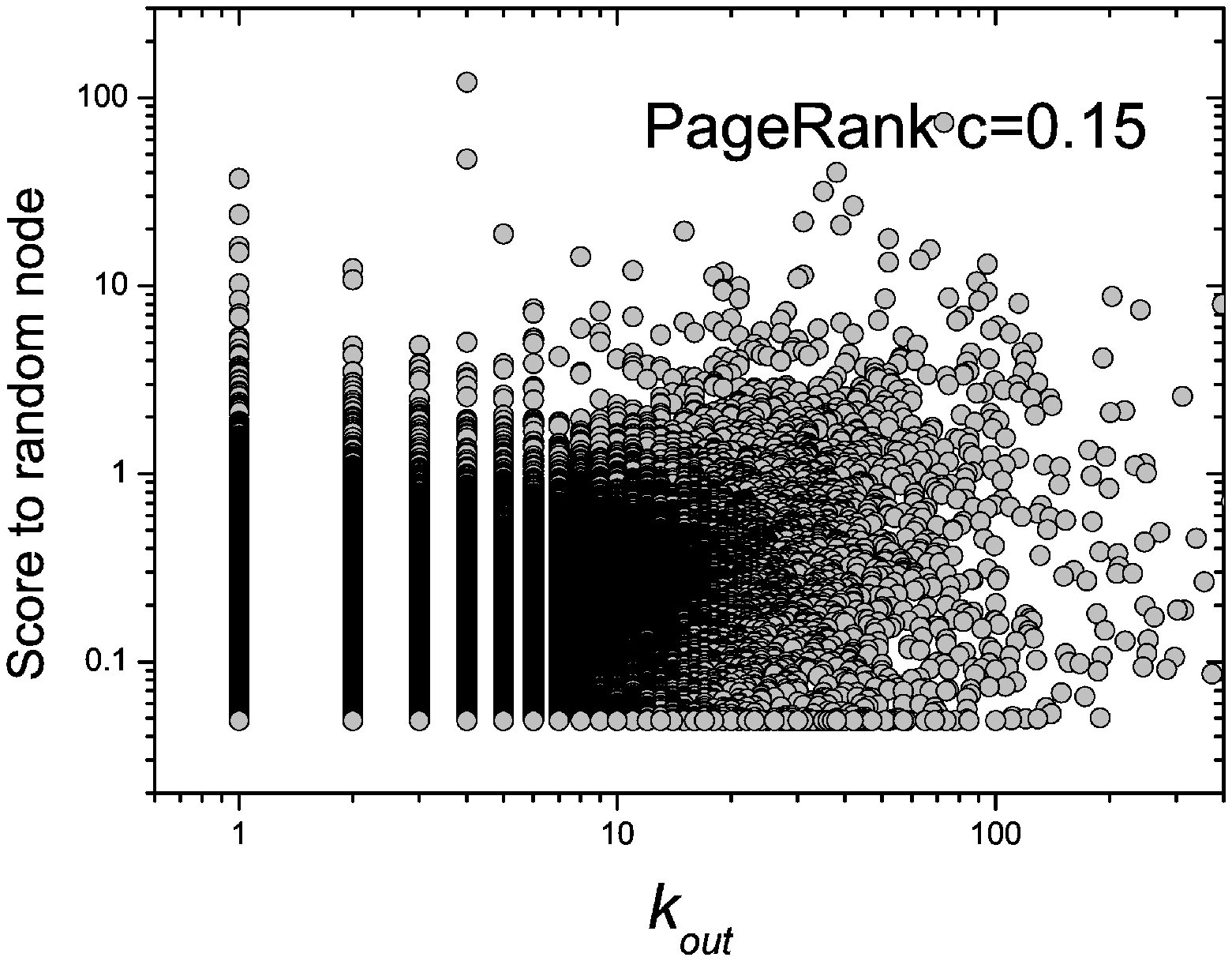, width=0.47\linewidth}}}
\caption{
The score flow from a node to (a) the ground node in LeaderRank and (b) random nodes in PageRank
as a function of $k_{\rm out}$, the number of leaders.
}
\label{fig_diff}
\end{figure}

The obvious difference between LeaderRank and PageRank lies in the formulation,
where the ground node in LeaderRank plays an important role in regulating probability flows,
making LeaderRank a parameter-free algorithm.
An essential difference does lie in the heart of dynamics.
In LeaderRank,
the score flow from node $i$ to the ground node is given by 
\begin{eqnarray}
	f_{i\rightarrow g}=\frac{s_i(t_c)}{k_i^{\rm out}},
\end{eqnarray}
while in PageRank the score flow from node $i$ to a random node is given by
\begin{eqnarray}
	f_{i\rightarrow {\rm rand}}=cs_i(t_c),
\end{eqnarray}
where $c$ is the return probability.
As shown in \fig{fig_diff},
$f_{i\rightarrow g}$ in LeaderRank is inversely proportional to the out-degree of $i$,
i.e. the number of leaders of $i$,
as expected from the above equation.
On the other hand,
$f_{i\rightarrow {\rm rand}}$ in PageRank show no obvious trend with the number of leaders.
Such observation corresponds to a fundamental difference between LeaderRank and PageRank.

\begin{figure}
\centerline{\epsfig{figure=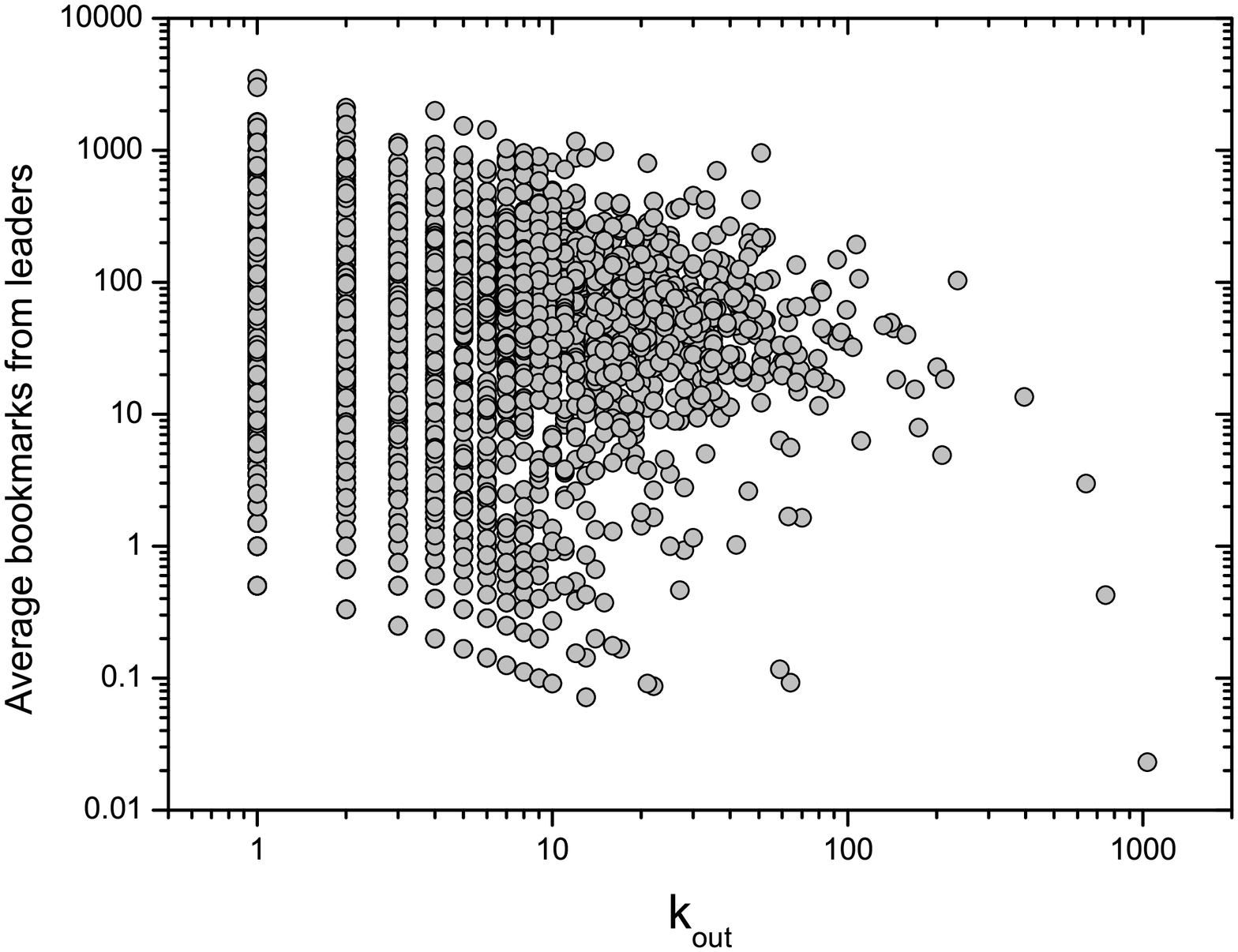, width=0.6\linewidth}}
\caption{
The ratio of saved bookmarks to the number of leaders as a function of $k_{\rm out}$.
}
\label{fig_avgbook}
\end{figure}

We may interpret the physical reasons in the following examples.
In social networks,
the score donated to the ground node can be interpreted as the information 
obtained from random browsing,
in contrast to the ordinary way of information acquisition from leaders.
The ground node can thus be considered as a centralized leader who provides general infomration.
We argue that
fans who have a large number of leaders may acquire less information from each leader,
including this centralized leader,
leading to the relation in \fig{fig_diff}(a).
Similar relation is observed in our empirical analyses with delicious data in \fig{fig_avgbook},
which show that the ratio of saved bookmarks to the number of leader,
decreases with  $k_{\rm out}$ of the user.
The same deduction can be obtained from the point of view of leaders.
If we assume that 
the average number of bookmarks provided by each leader is not indefinitely different,
nodes with small number of leaders receive only little information from leaders
and thus they have to acquire more information from the ground node.

In terms of ranking,
users with few leaders
should have small voting rights for leaders,
otherwise they may produce a strong bias if they donate all their score to only one or two leaders.
LeaderRank,
from which a negative correlation is introduced 
between score flow to leaders and out-degree
(i.e. flow to leaders is smaller from users with smaller out-degree),
would lead to a better ranking when compared to PageRank.

As the last example,
web surfers surfing on websites with small out-degree 
have limited choices of hyperlink
and by higher chance jump to another random website.
On the contrary,
web surfers are more likely to go through hyperlinks if there are lots of them on the website.
Such cases correspond to a 
small flow from nodes with large $k_{\rm out}$ to the ground node,
which is captured by LeaderRank.

%%%%%%%%% Figure 5 %%%%%%%%%%%%%%%%%
%%%%%%%%%%%%%%%%%%%%%%%%%%%%%%%%%%%%

%%%%%%%%% Figure 6 %%%%%%%%%%%%%%%%%
%%%%%%%%%%%%%%%%%%%%%%%%%%%%%%%%%%%%

\section{The top-100 ranked users}

Here we report the top 100 ranked users and their corresponding scores as obtained by LeaderRank, 
PageRank and the number of fans.
As one unit of score is initialized on every node in LeaderRank and PageRank,
the scores sum up to $N$ in these two rankings.
The last two columns show the top-100 users with the largest number of fans,
and their corresponding number of fans.

%%%%%%%%%%%%% Table top-20 users %%%%%%%%%%%%%%
\begin{longtable}{|c|c|c|c|c|c|c|}
\caption{Top 100 users ranked by LeaderRank, PageRank and the number of fans.} \label{tab_diffRank2} \\
%\rowcolors{4}{gray!20}{}
%\begin{tabular}{|c|c|c|c|c|c|c|}
\hline
Rank
& \multicolumn{2}{|c|}{LeaderRank} & \multicolumn{2}{|c|}{PageRank ($c$=0.15)} & \multicolumn{2}{|c|}{Number of fans} \\
\cline{2-7}
& User ID & Score & User ID & Score & User ID & Fans \# \\
\hline \hline
\endfirsthead
\hline
Rank
& \multicolumn{2}{|c|}{LeaderRank} & \multicolumn{2}{|c|}{PageRank ($c$=0.15)} & \multicolumn{2}{|c|}{Number of fans} \\
\cline{2-7}
& User ID & Score & User ID & Score & User ID & Fans \# \\
\hline \hline
\endhead
\hline
\endfoot
1 &  adobe  & 452 &  adobe  & 808 &  adobe  & 2768 \\
2 &  twit  & 382 &  twit  & 726 &  twit  & 2422 \\
3 &  wfryer  & 369 &  twitarmy  & 629 &  wfryer  & 1528 \\
4 &  willrich  & 358 &  thetechguy  & 536 &  willrich  & 1466 \\
5 &  joshua  & 264 &  cffcoach  & 529 &  merlinmann  & 1326 \\
6 &  cshirky  & 234 &  wfryer  & 492 &  joshua  & 1296 \\
7 &  hrheingold  & 217 &  willrich  & 475 &  steverubel  & 1284 \\
8 &  ewan.mcintosh  & 214 &  joshua  & 375 &  jgwalls  & 1142 \\
9 &  dwarlick  & 202 &  jdehaan  & 337 &  regine  & 1086 \\
10 &  twitarmy  & 200 &  lseymour  & 334 &  jonhicks  & 956 \\
11 &  merlinmann  & 186 &  isola  & 315 &  kevinrose  & 924 \\
12 &  blackbeltjones  & 171 &  cshirky  & 294 &  hrheingold  & 894 \\
13 &  jdehaan  & 170 &  secondlife  & 291 &  cshirky  & 837 \\
14 &  regine  & 170 &  ewan.mcintosh  & 288 &  dwarlick  & 827 \\
15 &  lseymour  & 168 &  hrheingold  & 285 &  zephoria  & 812 \\
16 &  jonhicks  & 168 &  merlinmann  & 267 &  ambermac  & 781 \\
17 &  zephoria  & 159 &  jonhicks  & 262 &  jgates513  & 702 \\
18 &  isola  & 159 &  samoore  & 261 &  ramitsethi  & 660 \\
19 &  djakes  & 158 &  dwarlick  & 261 &  ewan.mcintosh  & 635 \\
20 &  secondlife  & 156 &  kevinrose  & 256 &  cory$\_$arcangel  & 613 \\
21 &  edtechtalk  & 152 &  iwantsandy  & 249 &  secondlife  & 587 \\
22 &  steverubel  & 150 &  regine  & 248 &  brightideasguru  & 586 \\
23 &  jgwalls  & 142 &  jgwalls  & 234 &  judell  & 576 \\
24 &  kevinrose  & 135 &  steverubel  & 222 &  warrenellis  & 566 \\
25 &  brightideasguru  & 124 &  edtechtalk  & 214 &  edtechtalk  & 559 \\
26 &  jgates513  & 123 &  zephoria  & 212 &  elisebauer  & 545 \\
27 &  cogdog  & 120 &  nichoson  & 210 &  blackbeltjones  & 541 \\
28 &  joi$\_$ito  & 119 &  djakes  & 206 &  hokie62798  & 533 \\
29 &  cffcoach  & 114 &  blackbeltjones  & 206 &  djakes  & 531 \\
30 &  hokie62798  & 113 &  elisebauer  & 203 &  infosthetics  & 527 \\
31 &  samoore  & 112 &  dr.coop  & 178 &  bibliodyssey  & 509 \\
32 &  cityofsound  & 112 &  sdigrego  & 172 &  jakkarin  & 476 \\
33 &  heyjude  & 110 &  ambermac  & 161 &  chrisbrogan  & 474 \\
34 &  elisebauer  & 108 &  ureerat  & 160 &  russelldavies  & 461 \\
35 &  veen  & 104 &  jgates513  & 160 &  makemagazine  & 461 \\
36 &  shareski  & 102 &  glass  & 160 &  ericerb  & 455 \\
37 &  mathowie  & 101 &  brightideasguru  & 159 &  cityofsound  & 454 \\
38 &  thetechguy  & 101 &  ramitsethi  & 150 &  jummumboy  & 435 \\
39 &  judell  & 100 &  hokie62798  & 150 &  jdawg  & 433 \\
40 &  nichoson  & 100 &  cogdog  & 148 &  earlysound  & 430 \\
41 &  ambermac  & 99 &  joi$\_$ito  & 146 &  jzawodn  & 429 \\
42 &  warrenellis  & 96 &  heyjude  & 145 &  cogdog  & 428 \\
43 &  cory$\_$arcangel  & 93 &  judell  & 143 &  mathowie  & 421 \\
44 &  jutecht  & 92 &  cityofsound  & 142 &  plasticbag  & 407 \\
45 &  tomc  & 92 &  kawid  & 141 &  fredwilson  & 407 \\
46 &  choconancy  & 92 &  ceonyc  & 140 &  shanselman  & 406 \\
47 &  pedersoj  & 91 &  jdawg  & 139 &  heyjude  & 405 \\
48 &  mamamusings  & 91 &  bearsgonewild  & 136 &  leolaporte  & 404 \\
49 &  sdigrego  & 91 &  warrenellis  & 136 &  joi$\_$ito  & 385 \\
50 &  linkorama  & 90 &  benchaporn  & 134 &  samoore  & 384 \\
51 &  plasticbag  & 90 &  veen  & 130 &  curson12005  & 381 \\
52 &  sebpaquet  & 88 &  shareski  & 129 &  miyagawa  & 364 \\
53 &  ramitsethi  & 87 &  mathowie  & 127 &  veen  & 363 \\
54 &  snbeach50  & 83 &  choconancy  & 126 &  tuckermax  & 363 \\
55 &  ureerat  & 81 &  shanselman  & 126 &  kanter  & 359 \\
56 &  jdawg  & 81 &  jutecht  & 126 &  choconancy  & 354 \\
57 &  teach42  & 79 &  linkorama  & 124 &  deusx  & 351 \\
58 &  jakkarin  & 78 &  kick$\_$out$\_$the$\_$internet$\_$jams  & 123 &  aengle  & 351 \\
59 &  benchaporn  & 78 &  cory$\_$arcangel  & 123 &  lomo  & 350 \\
60 &  budtheteacher  & 77 &  selmav  & 121 &  bren  & 344 \\
61 &  infosthetics  & 75 &  pedersoj  & 119 &  wearehugh  & 342 \\
62 &  jzawodn  & 75 &  fju$\_$web20  & 114 &  53os  & 342 \\
63 &  raelity  & 73 &  mamamusings  & 113 &  101cookbooks  & 340 \\
64 &  chrisdodo  & 72 &  tomc  & 113 &  ginatrapani  & 336 \\
65 &  fredwilson  & 70 &  sebpaquet  & 111 &  angusf  & 333 \\
66 &  timo  & 70 &  bibliodyssey  & 111 &  zheng  & 331 \\
67 &  elemenous  & 69 &  apluscert  & 111 &  megsie  & 331 \\
68 &  bibliodyssey  & 69 &  alexdroege  & 109 &  britta  & 327 \\
69 &  iteachdigital  & 69 &  plasticbag  & 109 &  benchaporn  & 321 \\
70 &  timlauer  & 69 &  madro  & 108 &  teach42  & 319 \\
71 &  fstutzman  & 69 &  lisalis  & 108 &  knowhow  & 312 \\
72 &  foe  & 69 &  fredwilson  & 106 &  tomc  & 312 \\
73 &  migurski  & 69 &  infosthetics  & 105 &  snbeach50  & 307 \\
74 &  russelldavies  & 68 &  williams$\_$jeff  & 104 &  marisaolson  & 305 \\
75 &  alexdroege  & 67 &  101cookbooks  & 104 &  fstutzman  & 301 \\
76 &  curson12005  & 66 &  cablack  & 104 &  edans  & 300 \\
77 &  shanselman  & 65 &  snbeach50  & 103 &  jasonmcalacanis  & 298 \\
78 &  twitter$\_$edtech  & 65 &  jzawodn  & 103 &  williams$\_$jeff  & 292 \\
79 &  kick$\_$out$\_$the$\_$internet$\_$jams  & 64 &  wswu  & 103 &  yugop  & 290 \\
80 &  msippey  & 63 &  davepro14  & 102 &  wang1  & 290 \\
81 &  qdsouza  & 62 &  pamanapa  & 100 &  dhinchcliffe  & 288 \\
82 &  anne  & 62 &  fju$\_$webfund  & 100 &  ani625  & 288 \\
83 &  brasst  & 62 &  teach42  & 99 &  music  & 287 \\
84 &  aengle  & 61 &  tarisamatsumoto  & 98 &  elemenous  & 284 \\
85 &  ceonyc  & 61 &  fju$\_$univintro  & 96 &  toxi  & 282 \\
86 &  kfisch  & 61 &  russelldavies  & 95 &  google  & 281 \\
87 &  ehubbell  & 60 &  makemagazine  & 95 &  shareski  & 278 \\
88 &  makemagazine  & 60 &  fju$\_$inetcomp  & 95 &  mbauwens  & 275 \\
89 &  101cookbooks  & 59 &  clydekmann  & 93 &  design  & 275 \\
90 &  dr.coop  & 58 &  atrusty  & 92 &  mediaeater  & 274 \\
91 &  kanter  & 58 &  budtheteacher  & 92 &  ehubbell  & 271 \\
92 &  britta  & 58 &  elemenous  & 91 &  imao  & 270 \\
93 &  courosa  & 58 &  fstutzman  & 90 &  ureerat$\_$wat  & 267 \\
94 &  mguhlin  & 57 &  twitter$\_$edtech  & 90 &  ma.la  & 265 \\
95 &  marisaolson  & 56 &  curson12005  & 90 &  alexdroege  & 265 \\
96 &  williams$\_$jeff  & 56 &  timo  & 89 &  jewel$\_$lee27  & 264 \\
97 &  tuckermax  & 56 &  raelity  & 89 &  linkorama  & 262 \\
98 &  jummumboy  & 56 &  iteachdigital  & 89 &  raganwald  & 261 \\
99 &  district6  & 56 &  shiang  & 88 &  brasst  & 261 \\
100 &  chrislehmann  & 55 &  knowhow  & 87 &  budtheteacher  & 260 \\
\hline
%\end{tabular}
\end{longtable}
%%%%%%%%%%%%%%%%%%%%%%%%%%%%%%%%%%%%%%%%%%%%%

\section{Zipf's law}

\begin{figure}[h]
\centerline{\epsfig{figure=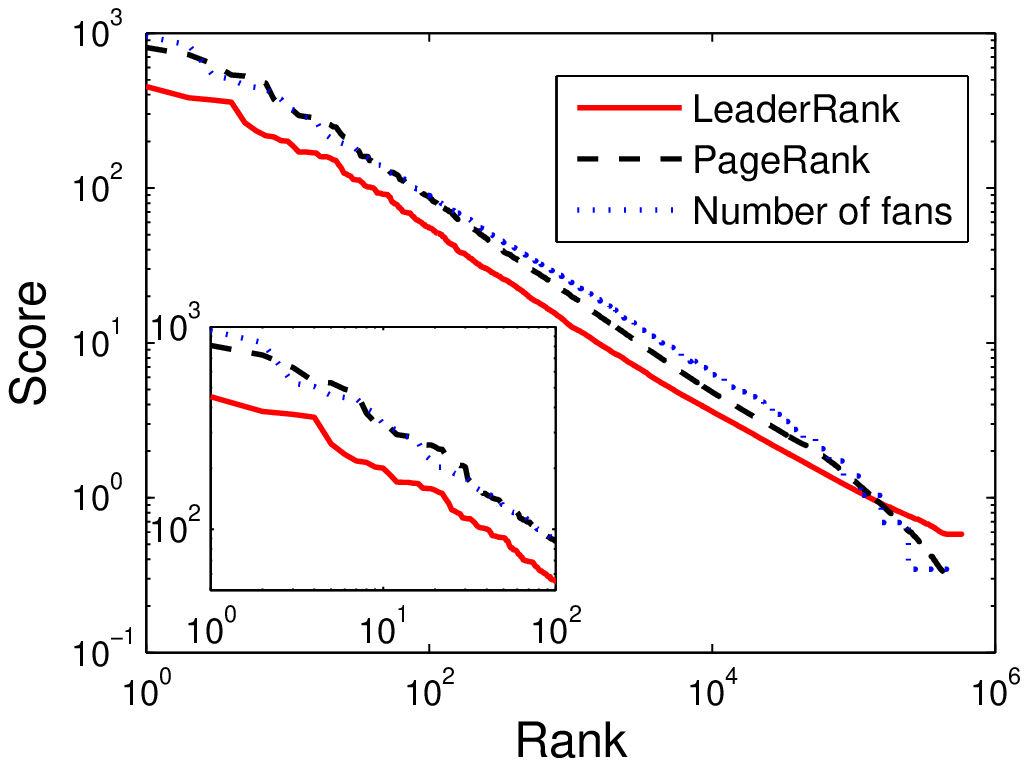, width=0.5\linewidth}}
\caption{
The score as a function of rank obtained from the LeaderRank, PageRank
and ranking by the number of fans.
Zipf's law is observed for these algorithms.
}
\label{fig_zipf}
\end{figure}

As shown in \fig{fig_zipf}, 
Zipf¡¦s law is observed for all the three ranking algorithms. 
We plot the score of each user against his/her rank and observe a power-law decaying. 
Notice that, 
although similar relation between score and rank is observed among the three algorithms, 
the ranking of individual is different by different algorithms

%%%%%%%%% Figure 6 %%%%%%%%%%%%%%%%%
%%%%%%%%%%%%%%%%%%%%%%%%%%%%%%%%%%%%

\section{Comparisons among ranking results from different ranking algorithms}

\begin{figure}
\centerline{\epsfig{figure=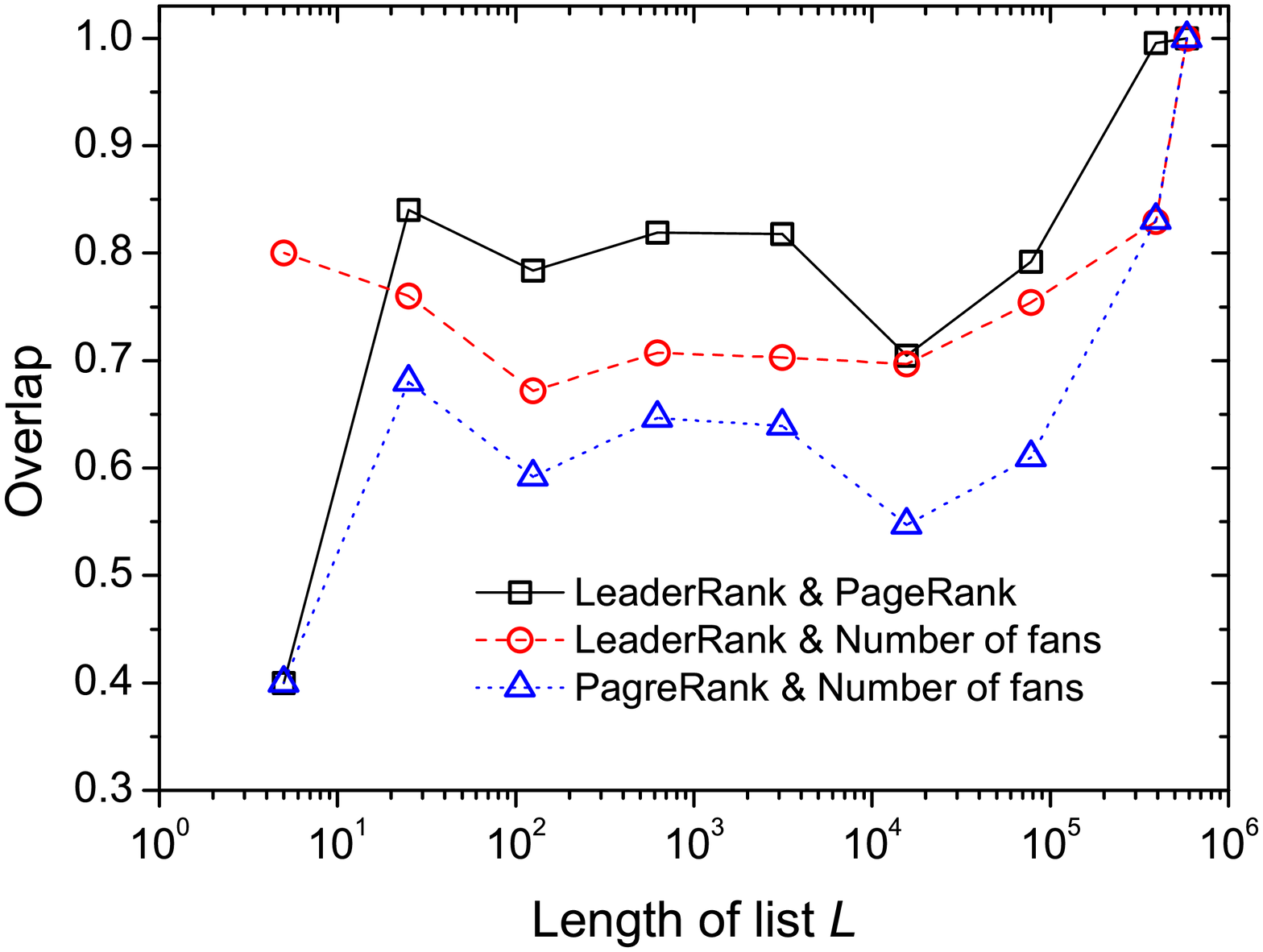, width=0.6\linewidth}}
\caption{
The overlap between LeaderRank and PageRank,
and LeaderRank and ranking by the number of fans,
as well as PageRank and ranking by the number of fans,
for the top-$L$ users.
}
\label{fig_overlap}
\end{figure}

\begin{figure}
\centerline{\epsfig{figure=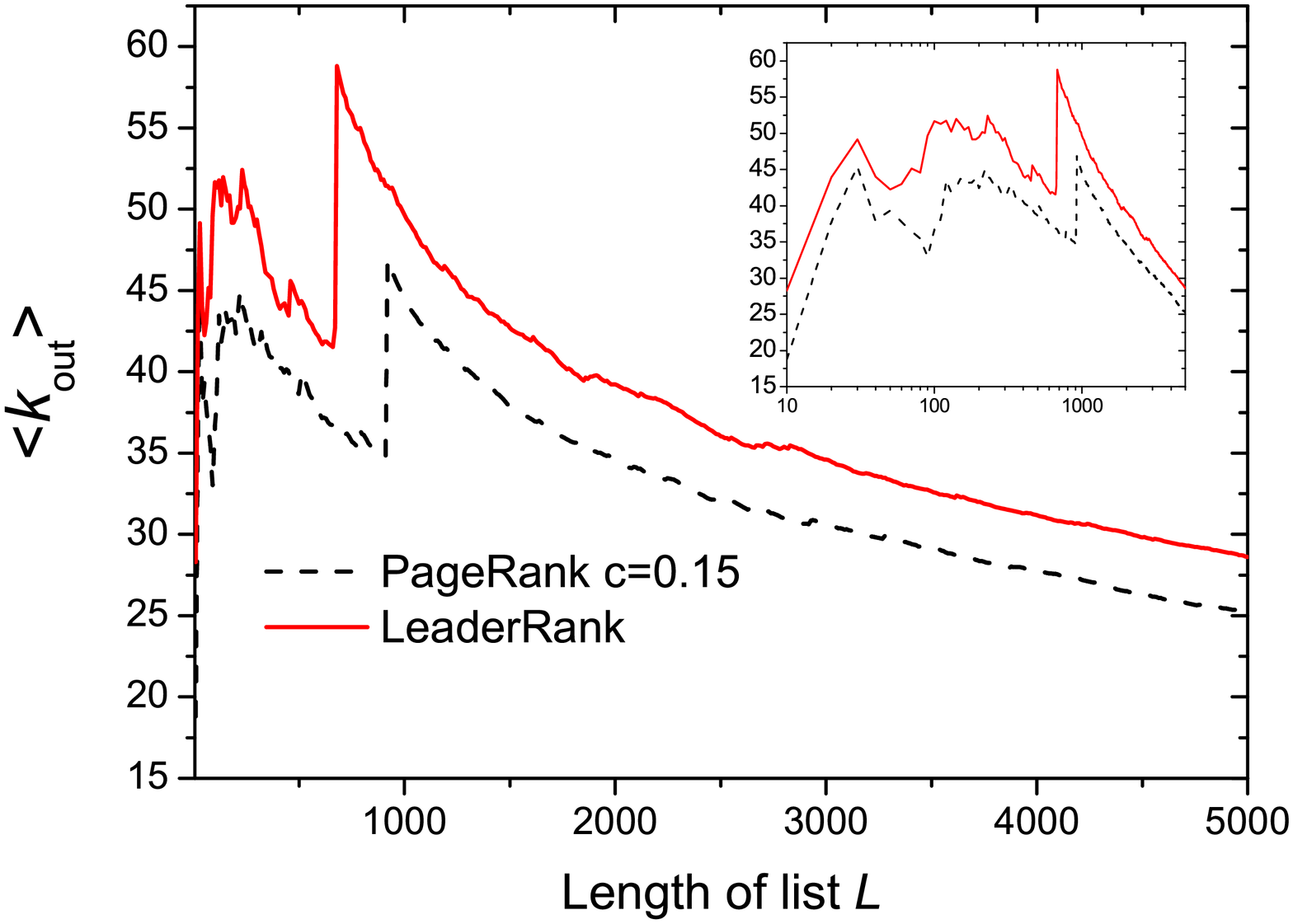, width=0.6\linewidth}}
\caption{
The average number of leaders of the top-$L$ users as ranked by LeaderRank and PageRank.
Inset: the average number of leaders against the logarithm of $L$. 
}
\label{fig_outDeg}
\end{figure}

We show in \fig{fig_overlap} the overlap of ranking between LeaderRank and PageRank,
as well as LeaderRank and the number of fans.
We plot as well the overlap between PageRank and the number of fans for reference.
These results show that LeaderRank is closer to PageRank,
than merely ranking by the number of fans,
and both LeaderRank and PageRank show positive correlation 
with the number of fans.
Though rankings from LeaderRank and PageRank seems to have large overlap,
the rankings of individual are different,
as can be seen in \tab{tab_diffRank}.
As shown in \fig{fig_outDeg},
average number of leaders of the top users as ranked by PageRank is always smaller than
that by LeaderRank.
It implies that
PageRank tends to assign high rank to nodes with small number of leaders,
which is unfair to nodes with large number of leaders.
We emphasize again individual rankings are different 
though the shape of the curves form LeaderRank and PageRank looks similar.

%%%%%%%%% Figure 6 %%%%%%%%%%%%%%%%%
%%%%%%%%%%%%%%%%%%%%%%%%%%%%%%%%%%%%

%%%%%%%%% Figure 6 %%%%%%%%%%%%%%%%%
%%%%%%%%%%%%%%%%%%%%%%%%%%%%%%%%%%%%

\section{Negative effect by removal of leaders}

\begin{figure}
\leftline{\epsfig{figure=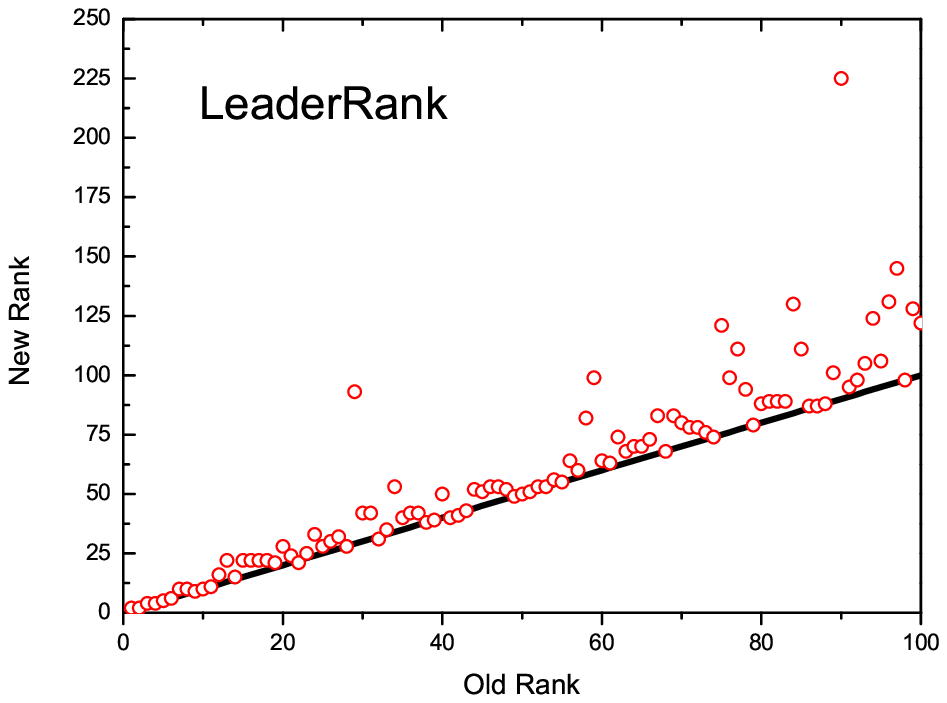, width=0.47\linewidth}                                                        
\leftline{\epsfig{figure=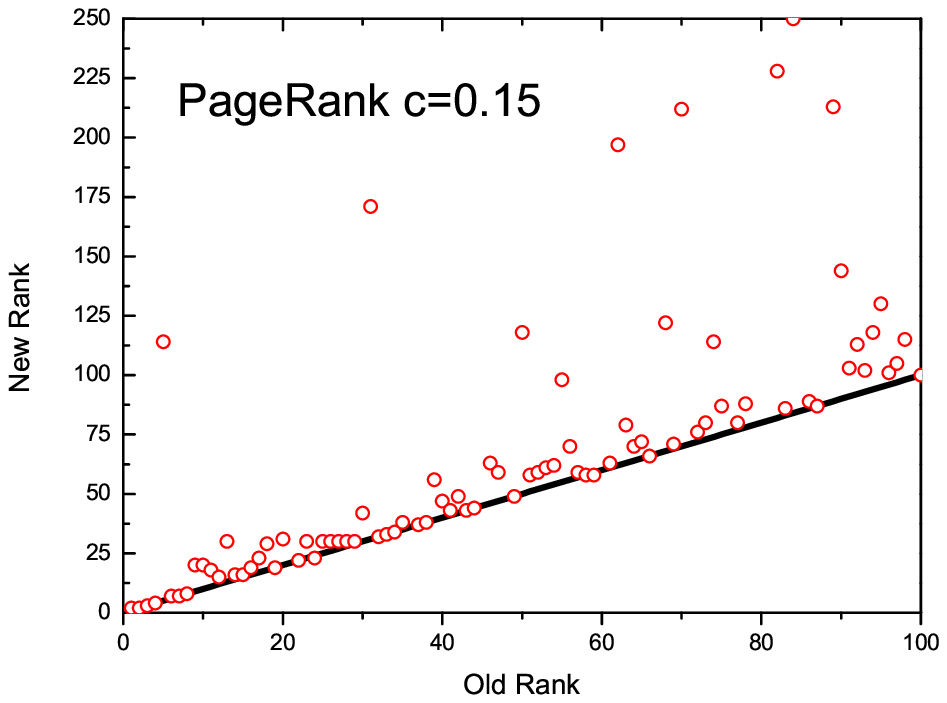, width=0.47\linewidth}}}
\caption{
The rank of a user after removing all his/her leaders,
as compared to his/her original rank as obtained by (a) LeaderRank and (b) PageRank.
The black solid line corresponds to the equality of the new and original rank.
}
\label{fig_remove}
\end{figure}

We show in \fig{fig_remove} that there is a negative effect in the rank of a user by removing
all his/her leaders.
As we can see for both LeaderRank and PageRank,
many users are lower in rank after removing their leaders.
These results suggest that considering just the leaders alone provides no absolute measure of influence,
as removing the entire upstream connection to leaders user may have a negative
effect on the social influence of an influential user.
In other words,
we have to consider the entire upstream topology to quantify the social influence of a user.

%%%%%%%%% Figure 5 %%%%%%%%%%%%%%%%%
%%%%%%%%%%%%%%%%%%%%%%%%%%%%%%%%%%%%

%%%%%%%%%%%%%%%%%%%%%%%%%%%%%%%%%%%%%%%%%%%%%%%%%%%%%%%%%%%%%%%%

\end{document}